\DocumentMetadata{}
\documentclass[sigconf]{acmart}
\AtBeginDocument{%
  \providecommand\BibTeX{{%
    \normalfont B\kern-0.5em{\scshape i\kern-0.25em b}\kern-0.8em\TeX}}}

\usepackage{listings}
\usepackage{multirow}
\usepackage{url}
\usepackage{graphicx}
\usepackage{subcaption} 
\usepackage{xcolor}
\usepackage{enumitem}
\usepackage{sc25repro}

\lstdefinelanguage{Mojo}{
 keywords={typeof, null, catch, switch, for, in, int, str, float, self, alias, Int, Float32, Float64, from, import, Scalar, UnsafePointer, print},
 keywordstyle=\color{blue}\bfseries,
 ndkeywords={boolean, throw, import},
 ndkeywords={return, class, if, elif, endif, while, do, else, True, False , catch, def, and, fn, var},
 identifierstyle=\color{black},
 sensitive=false,
 comment=[l]{\#},
 morecomment=[s]{/*}{*/},
 commentstyle=\color{purple}\ttfamily,
 stringstyle=\color{red}\ttfamily,
}

\setcopyright{cc}
\setcctype{by}
\acmConference[SC Workshops '25]{Workshops of the International Conference
for High Performance Computing, Networking, Storage and Analysis}{November
16--21, 2025}{St Louis, MO, USA}
\acmBooktitle{Workshops of the International Conference for High Performance
Computing, Networking, Storage and Analysis (SC Workshops '25), November
16--21, 2025, St Louis, MO, USA}
\acmDOI{10.1145/3731599.3767573}
\acmISBN{979-8-4007-1871-7/2025/11}

\definecolor{codegreen}{rgb}{0,0.6,0}
\definecolor{codegray}{rgb}{0.5,0.5,0.5}
\definecolor{codepurple}{rgb}{0.58,0,0.82}
\definecolor{backcolour}{rgb}{0.95,0.95,0.92}

\lstdefinestyle{mystyle}{
    backgroundcolor=\color{backcolour},
    commentstyle=\color{codegreen},
    keywordstyle=\color{magenta},
    numberstyle=\tiny\color{codegray},
    stringstyle=\color{codepurple},
    basicstyle=\ttfamily\footnotesize,
    breakatwhitespace=false,
    breaklines=true,
    captionpos=b,
    keepspaces=true,
    numbers=none,
    numbersep=5pt,
    showspaces=false,
    showstringspaces=false,
    showtabs=false,
    tabsize=2,
}

\lstdefinelanguage{Markdown}{
  sensitive=false,
  morecomment=[l]{\%},
}

\begin{document}

\title[Mojo HPC]{Mojo: MLIR-Based Performance-Portable HPC Science Kernels on GPUs for the Python Ecosystem}

\author{William F Godoy}
\authornote{Both authors contributed equally to this research.}
\orcid{0000-0002-2590-5178}
\affiliation{%
  \institution{Oak Ridge National Laboratory\thanks{This manuscript has been authored by UT-Battelle, LLC, under contract DE-AC05-00OR22725 with the US Department of Energy (DOE). The US government retains and the publisher, by accepting the article for publication, acknowledges that the US government retains a nonexclusive, paid-up, irrevocable, worldwide license to publish or reproduce the published form of this manuscript, or allow others to do so, for US government purposes. DOE will provide public access to these results of federally sponsored research in accordance with the DOE Public Access Plan (\url{https://www.energy.gov/doe-public-access-plan}).}}
  \city{Oak Ridge}
  \state{TN}
  \country{USA}
}
\email{godoywf@ornl.gov}

\author{Tatiana Melnichenko}
\authornotemark[1]
\orcid{0009-0001-0977-5481}
\affiliation{%
  \institution{Innovative Computing Laboratory. \\ The University of Tennessee, Knoxville.}
  \city{Knoxville}
  \state{TN}
  \country{USA}
}
\affiliation{%
  \institution{Oak Ridge National Laboratory.}
  \city{Oak Ridge}
  \state{TN}
  \country{USA}
}
\email{tdehoff@vols.utk.edu}

\author{Pedro Valero-Lara}
\orcid{0000-0002-1479-4310}
\affiliation{\institution{Oak Ridge National Laboratory}
\city{Oak Ridge}
\state{TN}
\country{USA} 
}
\email{valerolarap@ornl.gov}

\author{Wael Elwasif}
\orcid{0000-0003-0554-1036}
\affiliation{\institution{Oak Ridge National Laboratory}
\city{Oak Ridge}
\state{TN}
\country{USA} 
}
\email{elwasifwr@ornl.gov}

\author{Philip Fackler}
\orcid{0000-0003-4837-6181}
\affiliation{\institution{Oak Ridge National Laboratory}
\city{Oak Ridge}
\state{TN}
\country{USA} 
}
\email{facklerpw@ornl.gov}

\author{Rafael Ferreira Da Silva}
\orcid{0000-0002-1720-0928}
\affiliation{\institution{Oak Ridge National Laboratory}
\city{Oak Ridge}
\state{TN}
\country{USA} 
}
\email{silvarf@ornl.gov}

\author{Keita Teranishi}
\orcid{0000-0001-6647-2690}
\affiliation{\institution{Oak Ridge National Laboratory}
\city{Oak Ridge}
\state{TN}
\country{USA} 
}
\email{teranishik@ornl.gov}

\author{Jeffrey S Vetter}
\orcid{0000-0002-2449-6720}
\affiliation{\institution{Oak Ridge National Laboratory}
\city{Oak Ridge}
\state{TN}
\country{USA} 
}
\email{vetter@ornl.gov}

\renewcommand{\shortauthors}{Godoy, Melnichenko, et al.}

\begin{abstract}
We explore the performance and portability of the novel Mojo language for scientific computing workloads on GPUs. As the first language based on the LLVM's Multi-Level Intermediate Representation (MLIR) compiler infrastructure, Mojo aims to close performance and productivity gaps by combining Python's interoperability and CUDA-like syntax for compile-time portable GPU programming.
We target four scientific workloads: a seven-point stencil (memory-bound), BabelStream (memory-bound), miniBUDE (compute-bound), and Hartree--Fock (compute-bound with atomic operations); and compare their performance against vendor baselines on NVIDIA H100 and AMD MI300A GPUs.
We show that Mojo's performance is competitive with CUDA and HIP for memory-bound kernels, whereas gaps exist on AMD GPUs for atomic operations and for fast-math compute-bound kernels on both AMD and NVIDIA GPUs. Although the learning curve and programming requirements are still fairly low-level, Mojo can close significant gaps in the fragmented Python ecosystem in the convergence of scientific computing and AI.
\end{abstract}



\maketitle

\section{Introduction}
\label{sec:Introduction}

Mojo~\cite{Mojo} is the latest technology for closing performance and productivity gaps in the Python~\cite{van2007python} language and ecosystem, including in AI workloads.
Developed by Modular Inc., Mojo is a stand-alone language that provides ahead-of-time (AOT) and just-in-time (JIT) compilation options based entirely on the Multi-Level Intermediate Representation (MLIR)~\cite{lattner2019mlir}, which is the next-generation of the widely adopted LLVM~\cite{lattner2004llvm} compiler infrastructure. 
Mojo enables productivity through the reusability of Python's rich library ecosystem via module interoperability and offers performance portability across GPUs from different vendors.
It currently supports the latest NVIDIA GPUs, and support was recently added for AMD's MI300 GPUs. 
Hence, similar to modern programming languages such as Julia~\cite{Bezanson2017-ca} and Rust~\cite{matsakis2014rust}, Mojo aims to mitigate the costly fragmentation that arises when using multiple languages and ecosystems to achieve high-level characteristics such as productivity (e.g., Python) and safety combined with the low-level performance of compiled languages (e.g., C\texttt{++}, C, Fortran).

Heterogeneous computing architectures (e.g., CPUs + GPU accelerators) have been the primary drivers of the supercomputing and AI landscapes in recent decades.
The TOP500's\footnote{\url{www.top500.org}} Linpack~\cite{dongarra2003linpack} ranking from June 2025 shows that 9 out of the top 10 fastest supercomputers in the world are accelerated by GPUs. Moreover, massive investments in AI supercomputers (e.g., xAI's Colossus) continue to drive the demand for GPUs~\cite{pilz2025trendsaisupercomputers}.
For this reason, GPU programmability and portability are crucial for the entire high-performance computing (HPC) ecosystem.
Strategies to address portability needs have mainly targeted C, C\texttt{++}, and Fortran HPC languages through new standards, including OpenMP~\cite{openmp}, OpenCL~\cite{stone2010opencl}, OpenACC~\cite{wienke2012openacc}, and SYCL~\cite{deakin2020evaluating}, and through third-party programming models, including Kokkos~\cite{Kokkos3} and RAJA~\cite{Raja}. Since June of 2025, Mojo provides this capability of supporting portable GPU programming directly into the language standard library.

We attempt to answer the following research question: \textit{Can scientific users benefit from Mojo's performance-portable GPU codes?} Despite the Mojo language being particularly novel, with many design decisions still in the development stages, and not completely open-source until 2026, we provide an initial exploration of Mojo's unique value proposition---using it to write performance-portable GPU kernels using the language's standard library. 
Unlike AI methods and math kernels that are well defined in scientific numerical libraries, custom science kernels can have unique computational characteristics that affect their performance.
For this reason, we developed ports for four widely used scientific kernels---(i) seven-point stencil (memory-bandwidth bound), (ii) BabelStream (memory-bandwidth bound), (iii) miniBUDE (compute-bound), and (iv) Hartree--Fock (compute-bound with atomic operations)---so they can use Mojo's portable GPU code capabilities.
The kernels' performance and profiling metrics (where appropriate) are then analyzed and compared with their equivalent CUDA and HIP baselines. 

The rest of the paper is organized as follows: Section~\ref{sec:Background} provides background information on the Mojo language syntax and compile-time characteristics in connection with MLIR and an overview of the selected science kernels.
Section~\ref{sec:Methodology} introduces our Methodology for porting the codes to Mojo's memory and kernel-launching programming model and describes the targeted benchmarks and systems used in our experiments.
Section~\ref{sec:Results} describes the performance results and analysis of the Mojo-ported codes running on NVIDIA and AMD GPUs and compares the Mojo implementations with the corresponding vendor baselines in CUDA and HIP, respectively.
We also summarize our comparisons by applying a performance-portability metric to all of our runs on the NVIDIA and AMD GPUs.
Related work is presented in Section~\ref{sec:Related Work}, including attempts to close performance-portability and productivity gaps. Section~\ref{sec:Conclusions} presents the conclusions from our study. All software artifacts are provided for reproducibility in the Appendix section for Artifact Description and Evaluation.
To the best of our knowledge, this is the first work on Mojo's unifying portable GPU accelerator model to target scientific computing kernels.

\section{Background}
\label{sec:Background}

\subsection{Mojo Programming}

\begin{figure}[h]
  \centering
  \includegraphics[width=\columnwidth]{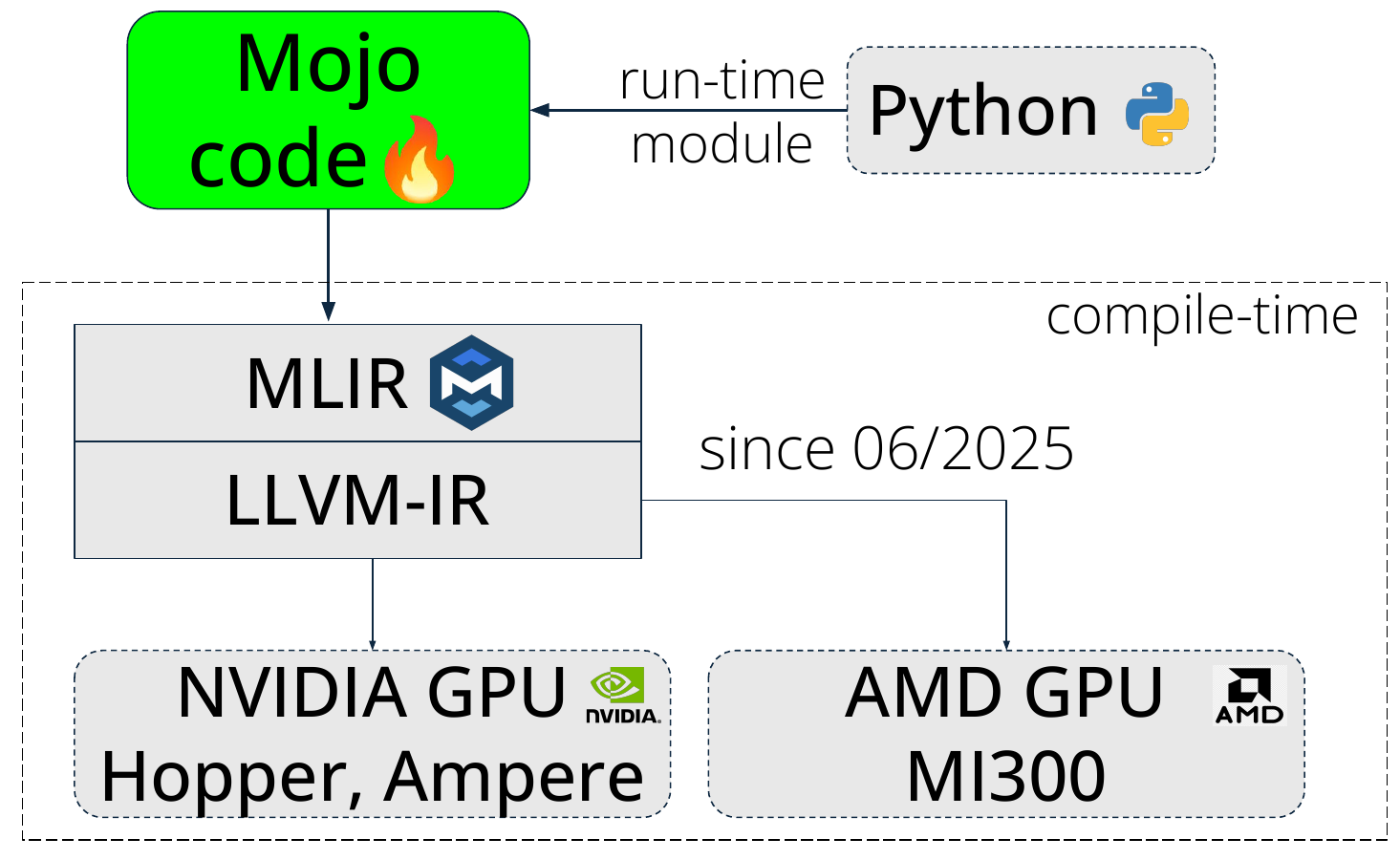}
  \caption{Mojo integrates Python at run time and performance-portable GPU code at compile time.}
  \label{fig:mojo}
\end{figure}

Mojo's value proposition is in combining the ease of use and flexibility of Python with the performance of compile-time programming languages such as C\texttt{++}. Mojo can achieve this by being the first language fully built on MLIR. Figure~\ref{fig:mojo} illustrates Mojo's approach to tackling the fragmentation between compiled languages and the Python ecosystem. 
We are particularly interested in how Mojo performs with science kernels given the June~2025 announcement of support for the AMD MI300 series GPUs in addition to the well-supported NVIDIA Hopper and Ampere GPUs.
Notably, reusing the same vendor-agnostic Mojo GPU modules adds a very unique characteristic for language-embedded GPU portability. This is in contrast with other HPC languages in which support is provided by standard third-party libraries. In fact, Mojo is unique in several ways:

\begin{itemize}
    \item Features MLIR-based compile time, JIT or AOT, for performance-portable GPU programming
    \item Includes run-time interoperability with Python and its rich library ecosystem
    \item Implements memory safety via variable lifetime and ownership (not garbage collected)
\end{itemize}

Listing~\ref{lst:Mojo} provides an example of Mojo's syntax, which illustrates these key features. Mojo uses a lower-level CUDA-/HIP-like programming model for GPU device memory allocation and kernel launching.
Because Mojo is intended to leverage compilation via MLIR, performance-critical information such as problem and block sizes, array layout, and variable types must be well-defined at compile time.
The latter also makes JIT or AOT indistinguishable for the generated intermediate representation because both mechanisms will be exposed to the same information.
This \textit{fixed layout} programming style was well suited for the early days of computing, but a shift in the dominant HPC languages has since occurred, in particular C and Fortran 90, and allowed for dynamic memory allocations, which are the building blocks of the majority of HPC codes.
In contrast, run-time interoperability with Python is done via modules, and it is not precompiled with MLIR, thus preserving Python's dynamic nature.
Hence, Mojo introduces this fragmentation in the language by keeping a clear separation between (i)~JIT- or AOT-MLIR compiled high-performance code and (ii)~Python interoperability as a run-time-only entity.

\begin{lstlisting}[label={lst:Mojo},caption={Compile-time GPU programming and run-time Python interoperability in Mojo.},language={Mojo},linewidth=0.9\linewidth,xleftmargin=.1\linewidth,basicstyle=\tiny]
# Mojo's gpu portable standard library module
from gpu.host import DeviceContext
from gpu.id import block_dim, block_idx, thread_idx
from layout import Layout, LayoutTensor
...
# Mojo's Python interoperability module
from python import Python

# Compile-time GPU programming
# Requires data tensor type, size and layout
alias dtype = DType.float32
alias Nx = 1024
alias layout = Layout.row_major(Nx)

alias block_size = 256
alias num_blocks = ceildiv(Nx, block_size)

# GPU kernel
fn fill_one(tensor: LayoutTensor[mut=True, dtype,layout]):
    var tid = block_idx.x * block_dim.x + thread_idx.x
    if tid < Nx:
        tensor[tid] = 1

fn main()
    # GPU memory model
    ctx = DeviceContext()
    d_u = ctx.enqueue_create_buffer[dtype](nx)
    u_tensor = LayoutTensor[dtype,layout](d_u)

    # GPU kernel execution
    ctx.enqueue_function[fill_one](u_tensor,
        grid_dim=num_blocks,
        block_dim=block_size
    )
    ctx.synchronize()

    # Python interoperability uses a separate runtime approach
    np = Python.import_module("numpy")
    array = np.array(Python.list(1, 2, 3))
    print(array)

\end{lstlisting}

Unlike most HPC software applications, in which much of the information is known only at run time (i.e., compile once, run many times), the reliance on MLIR optimizations restricts Mojo programs to providing performance-critical information (e.g., sizes, types, memory layouts, and variable lifetime) at compile time.
Although this is desired for JIT approaches (e.g., Julia), it can be a challenge when deploying code that has access to only certain information at run time (e.g., real-time workflows, adaptive mesh codes, statistical random walks).
Additionally, Python's stack is not compiled effectively with Mojo's AOT compilation (\texttt{mojo build}), thus keeping Python interactions dynamic and limited to run time through interoperability with CPython. 

Mojo also utilizes a memory model based on value ownership, which provides both performance and memory safety without relying on a garbage collector.
Heap-allocated values have exactly one owner at a time.
When a value's lifetime ends, Mojo automatically deallocates it by calling its destructor, thus eliminating the need for a garbage collector or the C\texttt{++} scope-based {\it resource allocation is initialization} pattern.

Mojo's approach to providing performance in the Python ecosystem is very different from other solutions. Rather than optimizing Python from within, Mojo makes a clear distinction between the rich compile-time performance-critical sections and the run-time Python interoperability.
While this is already the case for Python libraries based on C or C\texttt{++} thanks to performance-portable and interoperable libraries (e.g., pybind11, nanobind), Mojo's unique approach combines these aspects into the language to enhance productivity by minimizing fragmentation in Python-like software projects that require GPU performance and CUDA-like kernels.

Another important aspect is Mojo's interoperability with GPU vendor tools and ecosystems. In our early experience, debugging Mojo code is only supported by NVIDIA's Nsight Compute CLI (\texttt{ncu}) or the CUDA-GDB debugger (Mojo provides its LLDB-based debugger). Unfortunately, we were unable to find an officially supported tool to successfully profile with AMD's \texttt{rocprof}. Tooling is an important aspect of developing HPC software, and it is still an active area of development in Mojo.

\subsection{Science Kernels in Mojo}

Below is an overview of the selected science kernels and the implementations that we will use to assess Mojo's performance-portable capabilities on GPUs.  We focus on kernels from the following proxy applications:

\begin{itemize}
    \item \textbf{Seven-point stencil}: memory--bandwidth bound kernel in diffusion transport phenomena\footnote{\url{https://github.com/amd/amd-lab-notes}}
    \item \textbf{BabelStream}: memory transfer operations benchmarks ~\cite{deakin2018evaluating} 
    \item \textbf{miniBUDE}: in-silico molecular docking for protein prediction~\cite{poenaru2021performance} (compute-bound kernel)
    \item \textbf{Hartree--Fock}: quantum many-body approximation~\cite{osti_1783174} (compute-bound with atomics kernel)
\end{itemize}

\begin{figure}[ht]
\centering
\includegraphics[trim=0cm 1cm 0cm 1cm, width=\columnwidth]{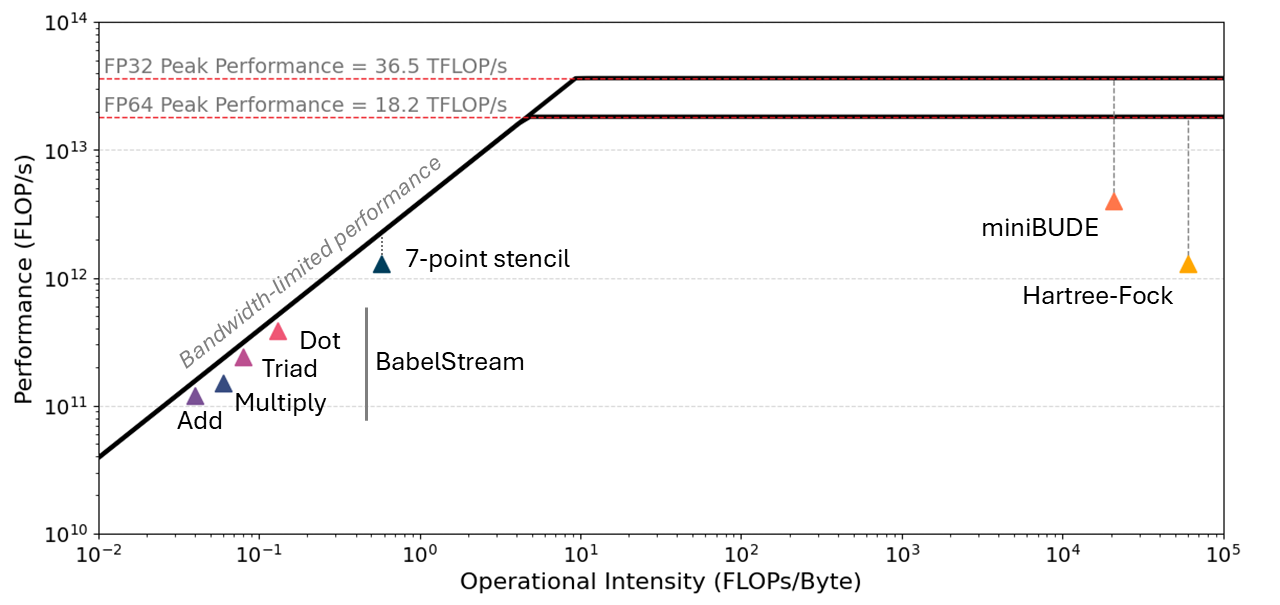}
\caption{{Roofline representation of the workloads implemented in the study using NVIDIA's CUDA and NSight on an NVIDIA H100 GPU.}}
\label{fig:roofline}
\end{figure}

These kernels provide initial coverage for memory and compute operations that are typical of codes used in many science domains. Figure~\ref{fig:roofline} presents a roofline model, obtained with the NVIDIA NSight profiler, in which these kernels lie in the memory- and compute-bound regions for the NVIDIA H100 GPUs. As shown, the model confirms the coverage obtained through these scientific kernels.
Because these kernels are well-established in HPC, CUDA and HIP C\texttt{++} codes are readily available and functional for comparison. In the rest of this section, we provide a brief overview of each proxy application. We explain the nature of each kernel and the computational characteristics that drive performance. We also provide snippets of the Mojo code that resulted from our porting tasks. 

\paragraph{Seven-point stencil} As a fundamental kernel that results from applying the Laplacian operator on partial differential equations (PDEs) for modeling diffusion phenomena, the seven-point stencil is used in several science domains.
Stencil calculations are an important class of PDE solvers on structured grids~\cite{5222004} and have been optimized for different hardware architectures, including GPUs~\cite{10.1145/2304576.2304619,KROTKIEWSKI2013533,doi:10.1080/10407790.2010.541359}.
The Mojo kernel implementation is presented in Listing~\ref{lst:7-point-stencil} and is a straightforward port from AMD's original HIP baseline implementation. The latter is also translated into CUDA by using the same structure as AMD's HIP code.

\begin{lstlisting}[label={lst:7-point-stencil},caption={Seven-point stencil portable Mojo GPU implementation.},language={Mojo},linewidth=0.99\linewidth,xleftmargin=.12\linewidth,basicstyle=\tiny]
alias precision = Float64
alias dtype = DType.float64
alias layout = Layout.row_major(L, L, L)
fn laplacian_kernel(f: LayoutTensor[mut=True, dtype, layout], 
                    u: LayoutTensor[mut=False, dtype, layout], 
                    nx: Int, ny: Int, nz: Int,
                    invhx2: precision, invhy2: precision, 
                    invhz2: precision, invhxyz2: precision
):
    var k = thread_idx.x + block_idx.x * block_dim.x
    var j = thread_idx.y + block_idx.y * block_dim.y
    var i = thread_idx.z + block_idx.z * block_dim.z

    if i > 0 and i < nx - 1 and j > 0 and j < ny - 1 and 
       k > 0 and k < nz - 1:
        f[i,j,k] = u[i,j,k] * invhxyz2
               + (u[i-1, j, k] + u[i+1, j, k]) * invhx2
               + (u[i, j-1, k] + u[i, j+1, k]) * invhy2
               + (u[i, j, k-1] + u[i, j, k+1]) * invhz2
\end{lstlisting}

\paragraph{BabelStream} As a simple implementation of the Stream benchmark for GPUs, BabelStream measures the memory-bandwidth limits of five common array kernel operations---\texttt{Copy}, \texttt{Multiply}, \texttt{Add}, \texttt{Triad}, and \texttt{Dot}---each of which is measured independently. 
Listing~\ref{lst:BabelStream} presents our initial efforts to port from the original CUDA and HIP codes to Mojo, highlighting how trivial this process can be for the first four operations because they allocate arrays exclusively on global memory. However, we expand on the use of shared block-memory capabilities in the \texttt{Dot} reduction operation.
BabelStream provides the building blocks of several memory--bandwidth bound algorithms (e.g., conjugate gradients) that are a composition of these basic operations.

\begin{lstlisting}[label={lst:BabelStream},caption={BabelStream portable Mojo GPU implementation.},language={Mojo},linewidth=0.9\linewidth,xleftmargin=.1\linewidth,basicstyle=\tiny]
fn copy_kernel(a: UnsafePointer[Scalar[dtype]], 
               c: UnsafePointer[Scalar[dtype]]):
    var i = block_dim.x * block_idx.x + thread_idx.x
    c[i] = a[i]

# add:      c[i] = a[i] + b[i]
# multiply: b[i] = scalar * c[i]
# triad:    a[i] = b[i] + scalar * c[i]

fn dot_kernel[size: Int](a:UnsafePointer[Scalar[dtype]], 
                         b:UnsafePointer[Scalar[dtype]], 
                         sums:UnsafePointer[Scalar[dtype]]):
    var tb_sum = stack_allocation[ TBSize, Scalar[dtype],
        address_space = AddressSpace.SHARED,
    ]()
    var i = block_dim.x * block_idx.x + thread_idx.x
    var local_tid = thread_idx.x

    threads_in_grid = block_dim.x * grid_dim.x
    while i < SIZE:
        tb_sum[local_tid] += a[i] * b[i]
        i += threads_in_grid

    var offset = block_dim.x // 2
    while offset > 0:
        barrier()
        if local_tid < offset:
            tb_sum[local_tid] += tb_sum[local_tid + offset]
        offset //= 2

    if local_tid == 0:
        sums[block_idx.x] = tb_sum[local_tid]
\end{lstlisting}

\paragraph{miniBUDE} As a proxy application to the Bristol University Docking Engine (BUDE) code, 
miniBUDE's core kernel, \texttt{fasten}, simulates the computational workloads involved in predicting the structure and strength of the interaction of two molecules.
This method is known as \textit{in-silico molecular docking} and is used for drug discovery.
The computational workload is driven by a few parameters: (i)~poses per work-item (PPWI), (ii)~work-group size, (iii)~number of ligands, and (iv)~number of proteins.
The Mojo implementation is shown in Listing~\ref{lst:miniBUDE} and captures the same structure as the original C\texttt{++} CUDA and HIP baseline codes.
We found that Mojo still lacks support for plain-old-data allocations on GPU memory for the original \texttt{Atom} struct consisting of 3-\texttt{Float32} and 1-\texttt{Int} elements.
Our workaround consists of mapping this struct into a flattened array of 4-\texttt{Float32} elements while casting back the last element as an \texttt{Int} inside the kernel.
Although not exactly the same as the CUDA and HIP versions, the Mojo code retains its portability.

\begin{lstlisting}[label={lst:miniBUDE},caption={miniBUDE Mojo kernel structure.},language={Mojo},linewidth=0.95\linewidth,xleftmargin=.05\linewidth,basicstyle=\tiny]
fn fasten_kernel[PPWI: Int](natlig: Int, natpro: Int,
                            protein_molecule: UnsafePointer[Float32],
                            ligand_molecule: UnsafePointer[Float32],
                            transforms_0: UnsafePointer[Float32],
                            ...
                            transforms_5: UnsafePointer[Float32],
                            etotals: UnsafePointer[Float32],
                            global_forcefield: UnsafePointer[Float32],
                            num_transforms: Int):
    
    var ix = block_idx.x * block_dim.x * PPWI + thread_idx.x
    if ix >= num_transforms:
        ix = num_transforms - PPWI

    var etot = SIMD[dtype, PPWI]()
    var transform = InlineArray[Vec4f32, PPWI * 3](uninitialized=True)

    var lsz = block_dim.x
    for i in range(PPWI):
        var index = ix + i * lsz
        sx: Float32 = sin(transforms_0[index]) ...
        transform[i * 3].x = cy * cz
    # Loop over ligand atoms
    while (True):
        ...
        for i in range(PPWI): ...
        # Loop over protein atoms
        while(True):
            for i in range(PPWI):
    # Write energy results
    var td_base = block_idx.x * block_dim.x * PPWI + thread_idx.x
    if td_base < num_transforms:
        for i in range(PPWI):
            etotals[td_base + i * block_dim.x] = etot[i] * Half
\end{lstlisting}

\paragraph{Hartree--Fock} This kernel provides a solution to the Hartree approximation of the Schr\"{o}dinger equation for a quantum many-body system at ground state. Wavefunctions are approximated by using a Slater determinant eliminating higher-order terms similar to density functional theory. The key kernel presented in Listing~\ref{lst:Hartree-Fock} performs dense symmetric matrix calculations of the electron-repulsion term of the Fock operator using integrals over Gaussian functions.
The kernel is composed of (i)~one parallelizable loop proportional to the fourth power of the number of atoms (\texttt{natoms}) with nested loops over the Gaussian components (usually three) and (ii)~six atomic operations over two square matrices of size \texttt{natoms}~$\times$~\texttt{natoms} representing the density and Fock terms.
Hence, the kernel is compute-bound with limited parallelism because of the atomic operations.
The Mojo, HIP, and CUDA implementations are based on the original Fortran~+~OpenMP implementation~\cite{osti_1783174}, and porting is a straight-forward process.

\newpage

\begin{lstlisting}[label={lst:Hartree-Fock},caption={Basic Hartree-Fock Fortran kernel structure.},language={Mojo},linewidth=0.95\linewidth,xleftmargin=0.05\linewidth,basicstyle=\tiny]
fn hartree_fock(ngauss: Int, schwarz: UnsafePointer[Float64],
                xpnt: UnsafePointer[Float64], 
                coef: UnsafePointer[Float64],
                geom: LayoutTensor[mut=True, dtype, geom_layout],
                dens: LayoutTensor[mut=True, dtype, layout],
                fock: LayoutTensor[mut=True, dtype, layout]):
  var ijkl = block_idx.x * block_dim.x + thread_idx.x
  # 4-levels nested ngauss loops
  for ib in range(ngauss): ...
    for jb in range(ngauss): ...
       for kb in range(ngauss): ...
         for lb in range(ngauss): ...
            eri += Float64(dkl * f0t * pow(aijkl, 0.5))
  # 6 atomic Fock matrix updates
  _ = Atomic.fetch_add(fock.ptr.offset(i * natoms + j),
                       rebind[Scalar[dtype]](dens[k, l] * eri * 4.0))
  ...
  _ = Atomic.fetch_add(fock.ptr.offset(j * natoms + l),
                       rebind[Scalar[dtype]](dens[i, k] * eri * -1))
\end{lstlisting}

\section{Methodology}
\label{sec:Methodology}

Our initial effort is a straightforward evaluation of the novel performance-portable capabilities of Mojo when running on GPUs.
Table~\ref{tab:Hardware} lists the hardware specifications of the NVIDIA~H100 and AMD~MI300A GPUs used in this study.

\begin{table}[h]
\caption{GPU Hardware Used in This Study}
\label{tab:Hardware}
\vspace{-0.5em}
\small
\centering
\resizebox{\linewidth}{!}{%
\begin{tabular}{@{}crrr@{}}
\toprule
GPU -- Memory   & \multicolumn{3}{c}{Theoretical Peaks}  \\
       &  Bandwidth      & FP32     & FP64      \\
       &  GB/s           & TFLOP/s & TFLOP/s  \\ 
\midrule
NVIDIA H100 NVL -- 94\,GB  &  3,900  &  60.0 & 30.0  \\
AMD MI300A -- 128 GB HBM3  &  5,300  & 122.6 & 61.3  \\
\bottomrule
\end{tabular}%
}
\end{table}

Each application is ported from a baseline implementation that uses the vendor-specific CUDA (NVIDIA) or HIP (AMD) programming models. Mojo's portable syntax shown in Listings~\ref{lst:7-point-stencil},~\ref{lst:BabelStream}, and ~\ref{lst:Hartree-Fock} is compared against available CUDA and HIP implementations. For each case, we use metrics that reflect a given application's purpose. 

\paragraph{Seven-point stencil} We use an effective memory bandwidth metric from the original baseline to represent the cell data advanced during a simulation step in GB/s. For simplicity, we use a constant $L$ size in each direction: $x$, $y$, and $z$. The metric is defined in Eq.~\ref{eqn:babelstream_bandwidth_eff} as a function of $L$ and the size of the element type ($T$ = Float32 or Float64) used in our runs. 

\begin{gather}
    fetch\_size_{effective} = \left [ L^3 - 8 - 12 \left (L - 2 \right ) \right ] \cdot sizeof(T) \label{eqn:fetch} \nonumber \\
    write\_size_{effective} = (L - 2)^3 \cdot sizeof(T) 
    \label{eqn:write}  \nonumber \\
    bandwidth_{effective} = \frac{ \left( fetch\_size + write\_size \right)_{effective} }{kernel\_time} 
\label{eqn:7p_bandwidth_eff}
\end{gather}

\paragraph{BabelStream} We use an effective memory bandwidth metric based solely on the size of the arrays (vector sizes) and the number of arrays for each operation.
This method is summarized in Eq.~\ref{eqn:babelstream_bandwidth_eff} for each of the fundamental operations.
Notably, although all operations allocate arrays on global device memory, the \texttt{Dot} operation exploits block-level shared memory for faster first-level reduction operations.

\begin{gather}
bandwidth_{Array} = \frac{sizeof(T) \cdot Vector\_size}{kernel\_time} \nonumber \\
bandwidth_{Copy} = 2 \cdot bandwidth_{Array} \nonumber \\
bandwidth_{Mul} = 2 \cdot bandwidth_{Array} \nonumber \\
bandwidth_{Add} = 3 \cdot bandwidth_{Array} \nonumber \\
bandwidth_{Triad} = 3 \cdot \cdot bandwidth_{Array} \nonumber \\
bandwidth_{Dot} = 2 \cdot \cdot bandwidth_{Array}
\label{eqn:babelstream_bandwidth_eff}
\end{gather}

\paragraph{miniBUDE} Given the compute-bound nature of miniBUDE's \texttt{fasten} kernel, the performance metric measures the floating-point operations per kernel execution time (GFLOP/s). We obtain this metric through a calculation from the original baseline:

\begin{gather}
ops_{workgroup} = 28\,PPWI + \nonumber \\ nligands \cdot \left [ 2 + 18\,PPWI + nproteins \cdot (10 + 30\, PPWI ) \right ] \nonumber \\
total_{ops} = ops_{workgroup} \frac{poses}{PPWI} \nonumber \\
miniBUDE-total_{gflops} = \frac{total_{ops}}{time_{kernel}} \cdot 1E-9
\label{eqn:minibude_gflops}
\end{gather}

\paragraph{Hartree--Fock} The figure of merit for this kernel is directly related to speedups obtained by porting the original algorithm~\cite{osti_1783174} to GPUs.  To compare across languages, we measure kernel wall-clock times directly without further transformations. We select available benchmarks for systems of helium atoms up to 256 atoms and 3 Gaussian functions per atom. As shown in Listing~\ref{lst:Hartree-Fock}, parallelization is heavily limited by the atomic operations inside this kernel. Our objective is to measure the impact of this pattern on Mojo's GPU-portable atomic operations capabilities.

For all the selected cases, we can establish that our methodology balances performance and portability for a fair assessment of the Mojo GPU code. Our methodology can be summarized as follows: 

\begin{itemize}
    \item Warm-up steps: For all codes, we discarded the first step in our measurements to reduce JIT or caching effects.
    \item Variability: We used dedicated hardware and collected several runs, at least 100, in our experiments to capture variability in our results. Kernel times have been verified with profiling tools.
    \item One-to-one comparisons: GPU code was not fine-tuned and matches any existing baseline configuration (e.g., number of blocks, threads for GPUs). For BabelStream, we use the optimized vendor-specific CUDA version, whereas our Mojo port follows a hybrid vendor-independent version to ensure portability and the best possible trade-offs on NVIDIA and AMD GPUs.
    \item Profiling: We profiled cases of interest in which Mojo can either surpass the vendor programming model or when differences are consistent.
    \item Results consistency: We used color codes for each programming model and a uniform figure of merit across all of the figures that present the results of application runs.
    \item JIT vs. AOT: Owing to the compile-time nature of Mojo, we did not observe meaningful differences between JIT and AOT kernel executions. Profiling of Mojo code with NVIDIA's Nsight CLI (\texttt{ncu}) and AMD's \texttt{rocprof} is only possible with AOT-compiled Mojo code.
\end{itemize}

\section{Results}
\label{sec:Results}

This section presents the performance results from our Mojo implementations and the available vendor-specific CUDA and HIP codes.

\begin{figure}[h]
\centering
\subfloat[H100]{\includegraphics[trim=0cm 0.5cm 0cm 1cm, width=\columnwidth]{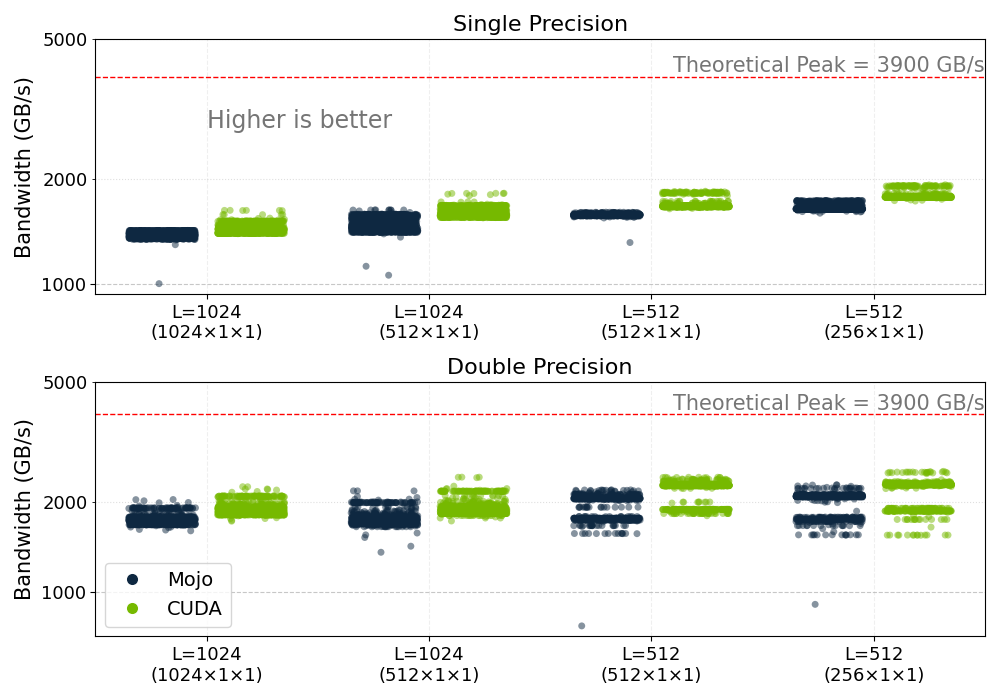}\label{fig:7p-CUDA}}\\
\vspace{0.3cm}
\subfloat[MI300A]{\includegraphics[trim=0cm 0.5cm 0cm 1cm, width=\columnwidth]{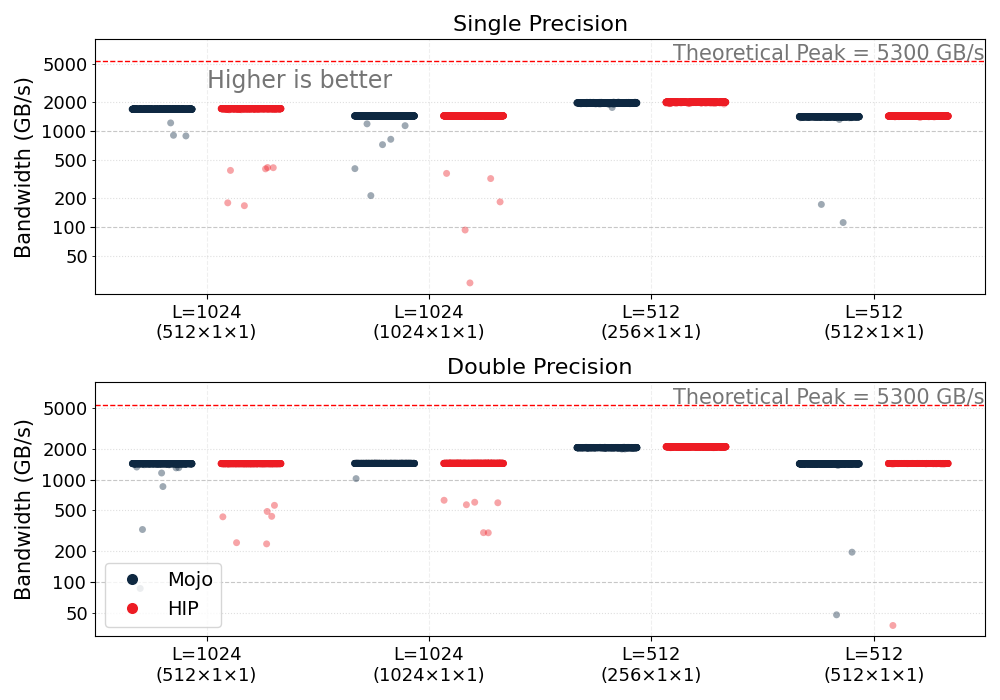}\label{fig:7p-HIP}}
\caption{{Mojo performance vs. CUDA and HIP for seven-point stencil kernels bandwidth.}}
\label{fig:results-7p}
\end{figure}

\paragraph{Seven-point stencil} Figure~\ref{fig:results-7p} shows raw data measurements from the obtained bandwidth (defined in Eq.~\ref{eqn:7p_bandwidth_eff}) and a comparison of Mojo versus CUDA  on an NVIDIA~H100~(\ref{fig:7p-CUDA}) and HIP on an AMD MI300A~(\ref{fig:7p-HIP}).
The code comparisons used two large problem sizes, 512 and 1024, while using a different number of blocks on the first grid dimension to achieve maximum performance.
Notably, tuning on the second or third dimension did not result in higher performance. 
As shown, Mojo is fairly competitive for both single and double precision, but is slightly slower on NVIDIA's H100, averaging 87\% of the CUDA performance.
This same variability is reflected in Mojo and the vendor-code runs on both GPUs but is more noticeable on the NVIDIA H100 when using double precision, hence our focus is on the overall differences across the raw data.
Results on the MI300A are essentially on par with the AMD HIP implementation, including capturing outlier measurements.
Interestingly, maximum bandwidth on the MI300A is reached with the same configuration, $L=512$ on a $512\times1\times1$ grid as recommended on the MI250X from the original HIP code.\footnote{\url{https://github.com/amd/amd-lab-notes}}

To further understand differences between Mojo and CUDA, we profiled the codes with NVIDIA's \texttt{ncu}.
Results are presented in Table~\ref{tab:7p-cuda-metrics} for the two cases shown in Fig.~\ref{fig:7p-CUDA}.
These cases were selected due to their more significant differences and overall slower performance. 
Overall, the CUDA-generated code exhibits more efficient use of memory resources, which is the most relevant aspect owing to the memory-bound nature of the kernel.
Performance differences arise from the use of registers at the local cache level; in fact, global memory loads and stores on both models are consistent with the expected 7 reads and 1 write for each cell value.
Moreover, Mojo kernels use more registers-per-thread (24) than CUDA (21) for the same operation, which explains the performance differences in L-caches and kernel duration.
Notably, the arithmetic intensity~(ai) and memory performance shown in the roofline analysis are consistent with our hypothesis.
Hence, there is room to further optimize the Mojo codes at the lower memory cache levels, even for a simple but widely used GPU stencil kernel.

\begin{table}[h]
\caption{Seven-Point Stencil Mojo vs. CUDA NCU Profiling Metrics}
\label{tab:7p-cuda-metrics}
\vspace{-0.5em}
\tiny
\centering
\resizebox{\linewidth}{!}{%
\begin{tabular}{lrr|rr}
\toprule
 & \multicolumn{2}{c|}{Double Precision} 
 & \multicolumn{2}{c}{Single Precision} \\
Nsight Compute CLI  & \multicolumn{2}{c|}{L=512 (512$\times$1$\times$1)}  & \multicolumn{2}{c}{L=1024 (1024$\times$1$\times$1)}  \\
(\texttt{ncu}) metric & Mojo & CUDA & Mojo  & CUDA \\
\midrule
Duration (ms)    & 1.10 & 0.96 & 8.74 & 7.21 \\
Throughputs (\%) &  &      & &  \\
  - Compute SM & 81.41 & 51.96 & 79.8 & 53.7 \\
  - Memory     & 67.98 & 76.72 & 37.7 & 43.9 \\
L1 ai (FLOP/byte)   &  \multicolumn{2}{c|}{0.14} & \multicolumn{2}{c}{0.24} \\ 
L2 ai (FLOP/byte)   &  \multicolumn{2}{c|}{0.26} & \multicolumn{2}{c}{0.51} \\ 
L3 ai (FLOP/byte)   &  \multicolumn{2}{c|}{0.62} & \multicolumn{2}{c}{1.24} \\ 
L1-3 Perf (FLOP/s)     & 1.20\,E12  & 1.38\,E12 & 1.22\,E12 & 1.48\,E12 \\ 
Registers & 24  & 21 & 26  & 20 \\
Load Global (LDG) & \multicolumn{2}{c|}{7} & \multicolumn{2}{c}{7}  \\
Store Global (STG) & \multicolumn{2}{c|}{1} & \multicolumn{2}{c}{1} \\
\bottomrule
\end{tabular}
}
\end{table}

\begin{figure}[ht]
\centering
\subfloat[H100]{\includegraphics[trim=0cm 0.5cm 0cm 0cm, width=\columnwidth]{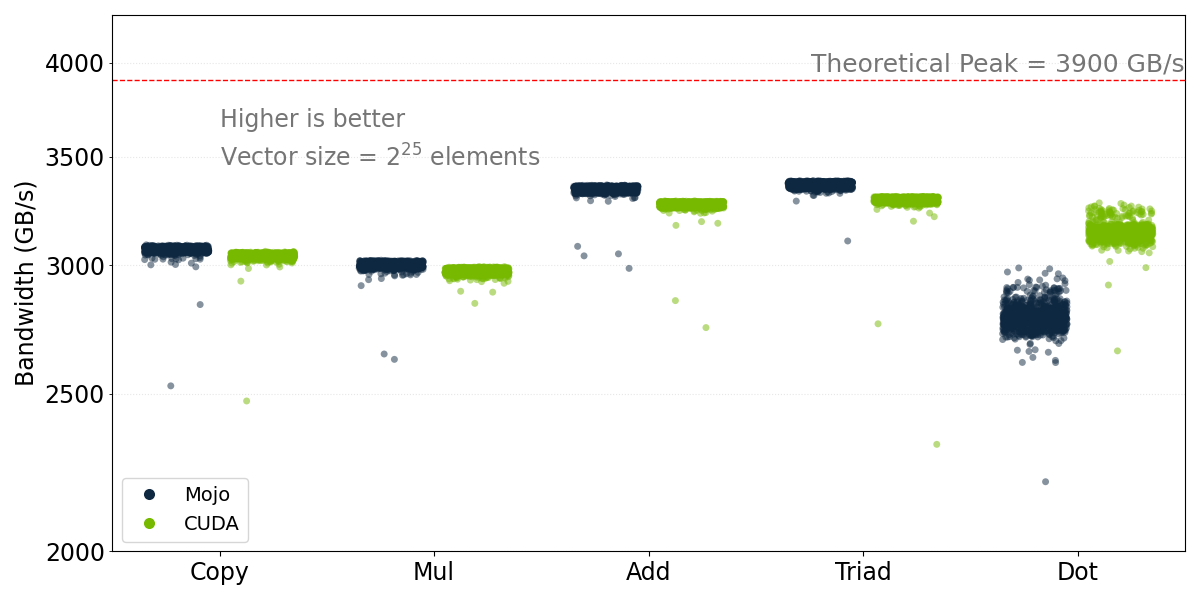}\label{fig:bs-CUDA}}\\
\vspace{0.3cm}
\subfloat[MI300A]{\includegraphics[trim=0cm 0.5cm 0cm 0cm, width=\columnwidth]{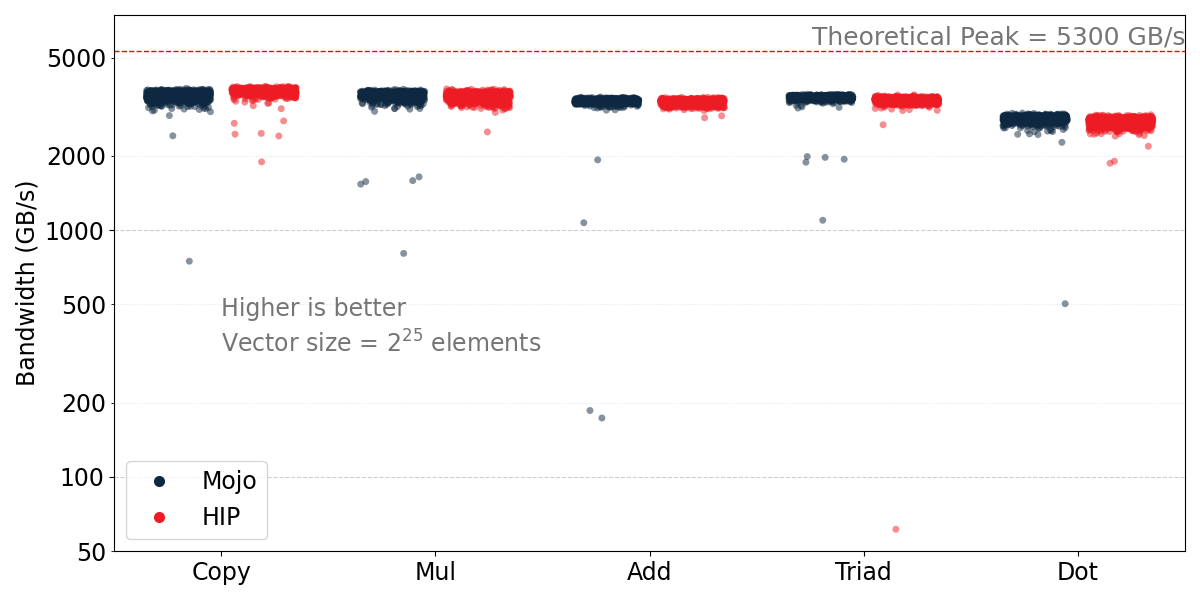}\label{fig:bs-HIP}}
\caption{{Mojo performance vs. CUDA and HIP for BabelStream's kernels bandwidth.}}
\label{fig:results-Babelstream}
\end{figure}

\paragraph{BabelStream} Figure~\ref{fig:results-Babelstream} shows the obtained bandwidth metrics from Eq.~\ref{eqn:babelstream_bandwidth_eff} for the fundamental \texttt{Copy}, \texttt{Mul}, \texttt{Add}, \texttt{Triad}, and \texttt{Dot} operations and compares the Mojo GPU portable code with vendor implementations for a large vector size of $2^{25} = 33,554,432$.
Surprisingly, except for the \texttt{Dot} operation, the Mojo implementation is slightly faster than CUDA, whereas Mojo's results on AMD are nearly indistinguishable from the HIP version.
This requires further analysis, but it is somewhat expected because the Mojo portable \texttt{Dot} product is the only operation that is not identical to the CUDA and HIP optimized versions involving block-level shared memory.
The baseline CUDA version provides heuristics to calculate block size for NVIDIA-specific GPUs based on multiprocessors counts.
We decided to keep the Mojo implementation portable and follow a hybrid approach between the CUDA and HIP versions to maximize performance portability.
All the codes are provided in the Appendix.
Unlike the seven-point stencil cases, BabelStream runs show less variability, which might be related to its simpler 1-to-1 memory-access that uses a 1D computational grid. 

We ran additional profiling analysis to understand the differences on the NVIDIA H100, as shown in Table~\ref{tab:babelstream-cuda-metrics}. We confirmed that the more efficient use of low-level resources leads to slightly higher memory throughput and lower compute throughput, which is desired in memory-bandwidth workloads.
Also, the kernel run characteristics, such as arithmetic intensity (FLOP/byte), performance (FLOP/s), number of registers per thread, and global memory activity, are consistent with the CUDA implementation.

\begin{table}[h]
\caption{BabelStream Mojo vs. CUDA NCU Profiling Metrics}
\label{tab:babelstream-cuda-metrics}
\vspace{-0.5em}
\small
\centering
\resizebox{\linewidth}{!}{%
\begin{tabular}{lrr|rr|rr|rr}
\toprule
Nsight Compute CLI  & \multicolumn{2}{c|}{Copy} & \multicolumn{2}{c|}{Mul}  & \multicolumn{2}{c|}{Add} & \multicolumn{2}{c}{Dot} \\
(\texttt{ncu}) metric & Mojo & CUDA & Mojo  & CUDA & Mojo  & CUDA & Mojo  & CUDA \\
\midrule
Duration (ms)       & 0.202   & 0.205   & 0.203   & 0.208 & 0.264  & 0.269 &  0.215 & 0.168 \\
Throughputs (\%)    &    &       &   &  &   &  &   & \\
  - Compute SM      & 16.3   & 28.6  & 18.2  & 28.2 & 15.9  & 27.3 & 51.1  & 11.4 \\
  - Memory          &  \textsc{69.7}  & 68.9  & 69.2  & 68.0 &  81.7 & 80.5 & 69.9 & 87.6 \\
L1 ai (FLOP/byte)   & \multicolumn{2}{c|}{--}   & \multicolumn{2}{c|}{0.06}  & \multicolumn{2}{c|}{0.04} & \multicolumn{2}{c}{0.13} \\ 
L2 ai (FLOP/byte)   & \multicolumn{2}{c|}{--}  & \multicolumn{2}{c|}{0.08} & \multicolumn{2}{c|}{0.05} & 0.14  & 0.13 \\ 
L3 ai (FLOP/byte)   & \multicolumn{2}{c|}{--} & \multicolumn{2}{c|}{0.12} & \multicolumn{2}{c|}{0.06} & 0.14  & 0.13 \\ 
L1-3 Perf (FLOP/s) &  \multicolumn{2}{c|}{--}   &  1.64\,E11  & 1.61\,E11 & 1.26\,E11  & 1.24\,E11 & 3.5\,E11  & 4.01\,E11 \\ 
Registers          & \multicolumn{2}{c|}{16}  &  \multicolumn{2}{c|}{16} & \multicolumn{2}{c|}{16} & 26  & 20 \\
Load Global (LDG) & \multicolumn{2}{c|}{1}  &  \multicolumn{2}{c|}{1}  &  \multicolumn{2}{c|}{2}  &  \multicolumn{2}{c}{2} \\
Store Global (STG) & \multicolumn{2}{c|}{1}  &  \multicolumn{2}{c|}{1}  & \multicolumn{2}{c|}{1}  &   \multicolumn{2}{c}{1} \\
\bottomrule
\end{tabular}
}
\end{table}

\begin{figure*}
\centering
\includegraphics[width=0.98\linewidth,height=0.43\linewidth]{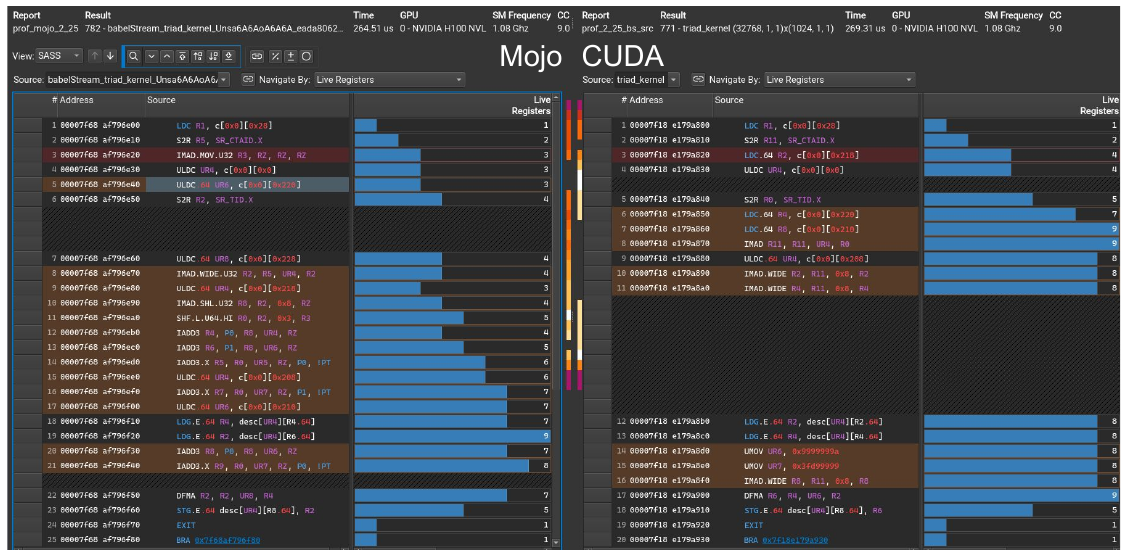}
\caption{{Mojo vs. CUDA generated assembly (SASS) from NVIDIA's Nsight for BabelStream's \texttt{Triad} kernel.}}
\label{fig:sass-triad-babelstream}
\end{figure*}

Figure~\ref{fig:sass-triad-babelstream} shows a deeper side-by-side assembly-level analysis that uses NVIDIA's Nsight tool for the \texttt{Triad} kernel.
This kernel, which is not shown in Table~\ref{tab:babelstream-cuda-metrics}, exhibits performance similar to the \texttt{Add} kernel, as expected.
We made three important observations:
(i)~Mojo produces fewer load constant operations than CUDA and uses constant memory without further annotations in the code,
(ii)~Mojo reports fewer active live registers but reports more compute operations (\texttt{IADD3}) during the kernel's main operation,
and (iii)~global memory loads (LDG) and global stores (STG) are consistent among codes.
The presented analysis is enough to understand the performance and throughput differences, and any further analysis would be beyond the scope of our performance-portability focus.

\begin{figure}[]
\centering
\subfloat[wg8]{\includegraphics[trim=1.5cm 0.5cm 1.5cm 0.5cm, width=\columnwidth]{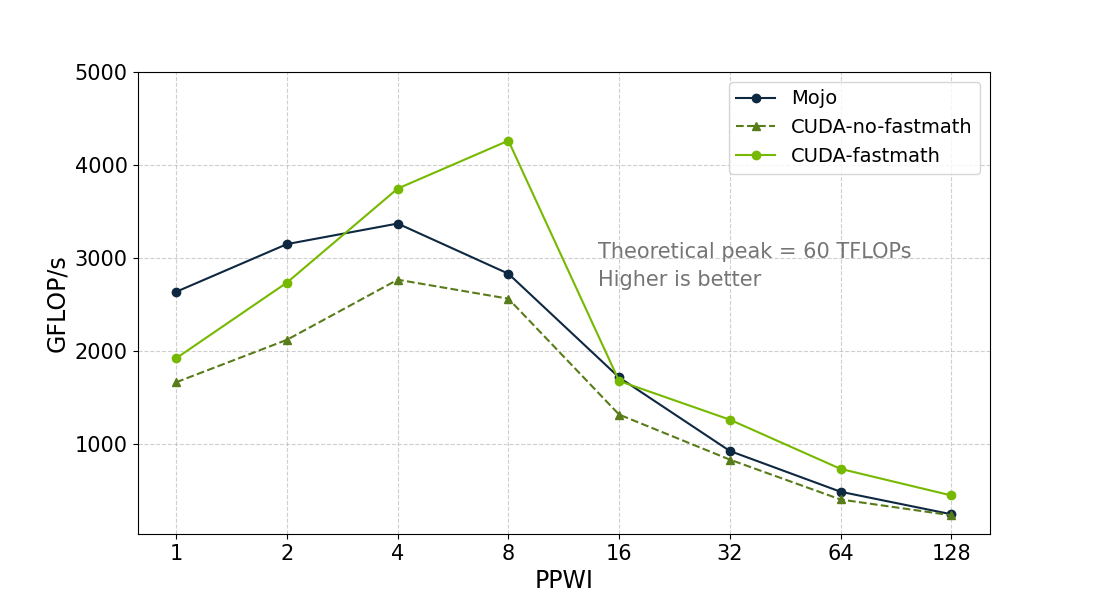}\label{fig:mini8-CUDA}}\\
\subfloat[wg64]{\includegraphics[trim=1.5cm 0.5cm 1.5cm 0cm, width=\columnwidth]{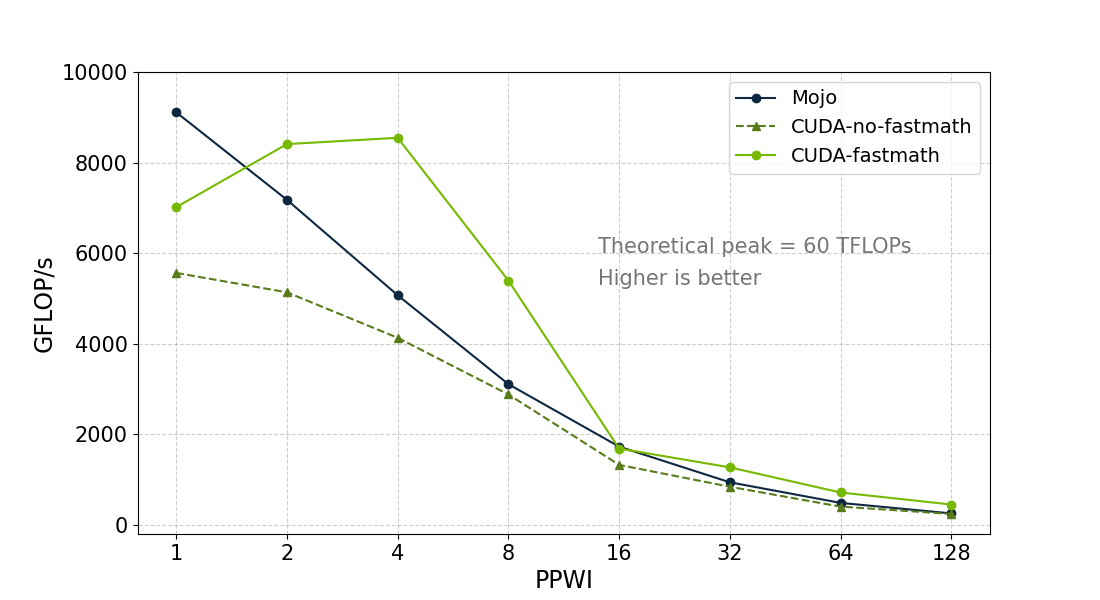}\label{fig:mini64-CUDA}}
\caption{{Mojo performance vs. CUDA for miniBUDE kernels on NVIDIA H100.}}
\label{fig:results-CUDA-minibude}
\end{figure}

\begin{figure}[h]
\centering
\subfloat[wg8]{\includegraphics[trim=1.5cm 0.5cm 1.5cm 0.5cm, width=\columnwidth]{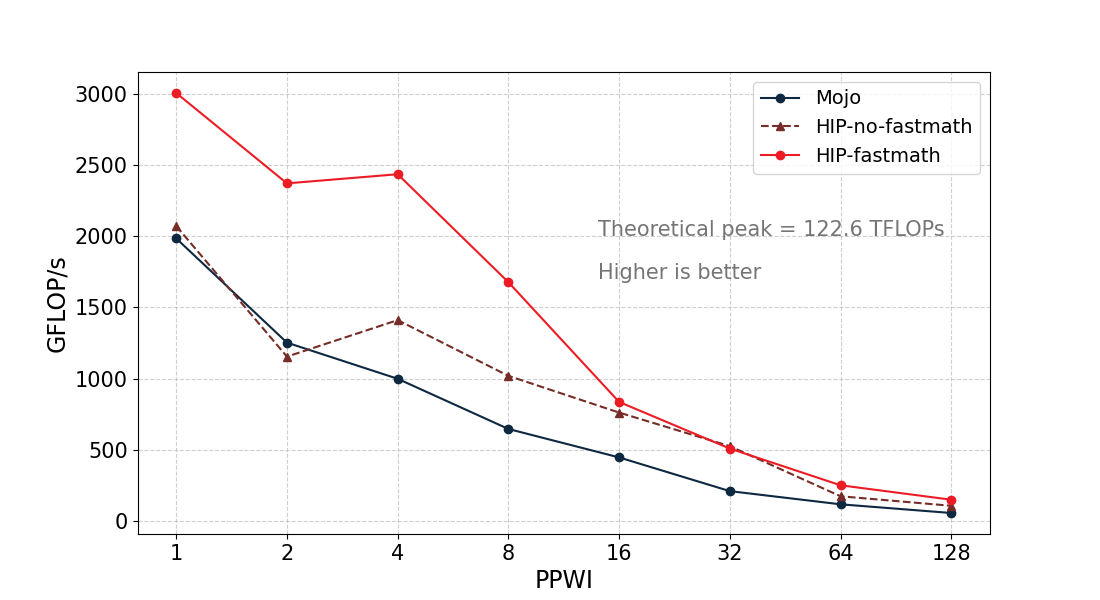}\label{fig:mini8-AMD}}\\
\subfloat[wg64]{\includegraphics[trim=1.5cm 0.5cm 1.5cm 0cm, width=\columnwidth]{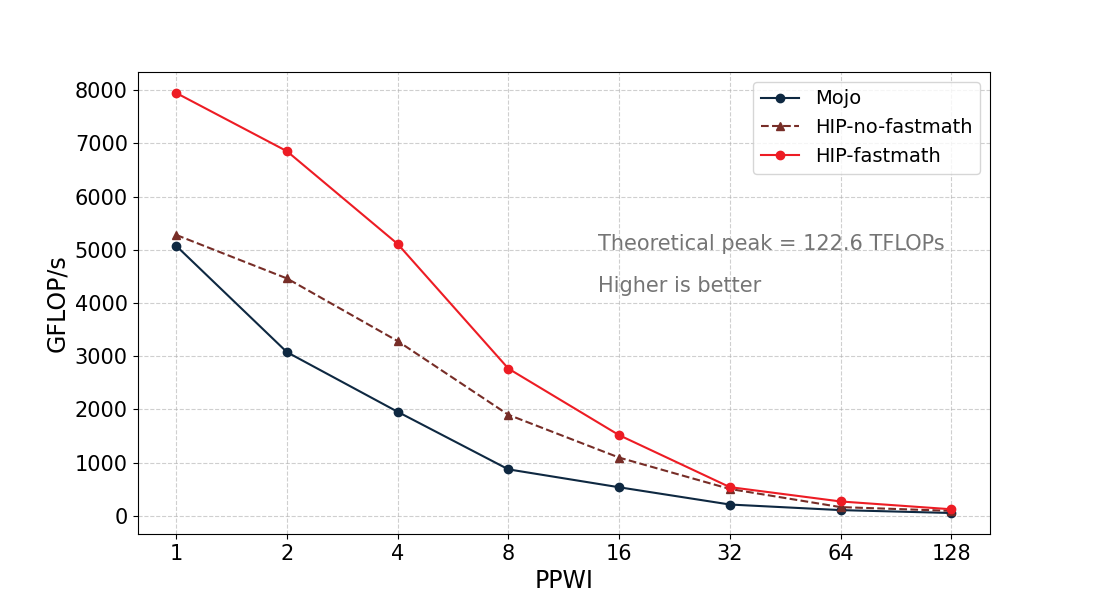}\label{fig:mini64-AMD}}
\caption{{Mojo performance vs. HIP for miniBUDE kernels on AMD MI300A.}}
\label{fig:results-AMD-minibude}
\end{figure}

\paragraph{miniBUDE} Figures~\ref{fig:results-CUDA-minibude} and \ref{fig:results-AMD-minibude} show the performance obtained in GFLOP/s from Eq.~\ref{eqn:minibude_gflops} in our experiments for different PPWI sizes and two work-group sizes (wg) (8 and 64) to vary the computational workload of the small bm1 benchmark.
We observe less variability for all cases, so we show only the average value obtained from our runs.
We also present CUDA and HIP runs without \texttt{fast-math} because this compute-bound workload is sensitive to such optimizations, and Mojo does not provide this option.
Mojo did not reach the same performance levels as it did in previous cases, but it follows a similar trend.
Mojo outperforms CUDA for a small PPWI and wg and is somewhere in between the fast-math optimized and non-optimized versions of CUDA on the NVIDIA H100.
The same Mojo code underperforms both the HIP fast-math optimized and non-optimized versions on the AMD MI300A, similar to the small PPWI and wg case shown in Figure~\ref{fig:mini8-AMD}.

\paragraph{Hartree--Fock} Our experience confirms the impact of the atomic operations preventing more parallelization.
Because of this noticeable effect, we provide only the wall-clock times as a function of the number of atoms, \texttt{a}, and Gauss functions per atom, \texttt{ngauss}, when comparing Mojo with CUDA and HIP (Table~\ref{tab:hartree-fock-metrics}).
Surprisingly, Mojo is 2.5$\times$ faster than CUDA on operations of up to 256 in size, but performance decreases dramatically for the 1024 operation.
Rather than drawing general conclusions, we believe that further analysis is necessary to understand these differences.
The opposite happened on AMD GPUs: Mojo largely underperforms the HIP implementation in all cases, which leads us to believe that we are still hitting a corner case in the language. 
Thus, as Mojo continues to mature on AMD GPUs, these capabilities are expected to improve at the lower levels.

\begin{table}[h]
\caption{Hartree--Fock Mojo vs. CUDA and HIP Metrics}
\label{tab:hartree-fock-metrics}
\vspace{-0.5em}
\tiny
\centering
\resizebox{\linewidth}{!}{%
\begin{tabular}{lrr|rr}
\toprule
Kernel execution      & \multicolumn{2}{c|}{NVIDIA H100} & \multicolumn{2}{c}{AMD MI300A} \\
duration (ms) & Mojo & CUDA & Mojo  & HIP \\
\midrule
a=1024 ngauss=6 &  147,250 &  2,652  & --     & 846 \\
a=256 ngauss=3  &  187     &   472  & 25,266  & 178 \\
a=128 ngauss=3  &   21     &    53  &  2,765  &  23 \\
a=64 ngauss=3   &    3     &     7  &    436  &   4 \\
\bottomrule
\end{tabular}
}
\end{table}

\subsection{Performance Portability Metric}

Given the results across the two selected GPU platforms, we apply a performance portable-metric ($\overline{\Phi}$) that was discussed in several recent publications~\cite{MAROWKA2025107826,9407113,pennycook2021revisiting,pennycook2019implications}.
We use the definition of $\overline{\Phi}$ for ``application efficiency'' as {\it the arithmetic mean of an application's performance efficiency observed across a set of platforms from the same architecture class}.
Thus, the metric is defined as the arithmetic mean of the best possible result:

\begin{gather}
    \overline{\Phi}_{Mojo} = \frac{\sum_{i \in T} e_i(a)}{\mid T \mid}        \\
    e_i(a) = \frac{Mojo_{perf-i}}{vendor(CUDA/HIP)_{perf-i}}, \nonumber
    \label{eqn:PerformancePortability}
\end{gather}

\noindent where $e$ is the efficiency of the Mojo performance defined as the ratio of the Mojo performance metric to the vendor-specific CUDA and HIP implementations. $T$ is the subset of runs for a particular proxy and architecture class (e.g., GPUs only). The results from applying this metric are shown in Table~\ref{tab:performance-portability}. 

\begin{table}[h]
\caption{Mojo Performance Portability Metric}
\label{tab:performance-portability}
\vspace{-0.5em}
\small
\centering
\resizebox{0.7\linewidth}{!}{%
\begin{tabular}{lc|c}
\toprule
Mojo        & NVIDIA  & AMD       \\
Efficiency  & H100    & MI300A \\
\midrule
7-point stencil & & \\
FP32 &   0.82  &  1.00  \\
FP64 &   0.87  &  1.00  \\
$\overline{\Phi}$ = 0.92        &         &        \\
\midrule
BabelStream & & \\
Copy     &   1.01   &  1.00  \\
Mul      &   1.02   &  1.00  \\
Add      &   1.01   &  1.00  \\
Triad    &   1.01   &  1.00  \\
Dot      &   0.78   &  1.00  \\
$\overline{\Phi}$ = 0.96  &    &        \\
\midrule       
miniBUDE &    &        \\
PPWI=8 wg=8   &   0.82   &  0.38  \\
PPWI=4 wg=64  &   0.59   &  0.38  \\
$\overline{\Phi}$ = 0.54 &          &        \\
\midrule
Hartree--Fock & & \\
a=1024 ngauss=6  &   17E-3  &  -- \\
a=256 ngauss=3   &   2.52   &  7E-3  \\
a=128 ngauss=3   &   2.52   &  8E-3 \\  
a=64 ngauss=3   &    2.33   & 8E-3 \\
$\overline{\Phi}$ = 0.92 & & \\
\bottomrule
\end{tabular}
}
\end{table}

The most important takeaway is that Mojo scores very high for seven-point stencil and BabelStream memory-bandwidth bound workloads.
Gaps exist in the compute-bound miniBUDE for both vendors, and these gaps are associated with the need for further optimization (e.g., fast-math).
Hartree--Fock results come from a special case with atomic operations in which Mojo outperforms CUDA but is far from achieving HIP's performance.
Additionally, the high value of the proposed performance methodology for the latter case can be misleading because it is the product of canceling out the outperforming of one vendor with the underperforming of another.

\subsection{Discussion and Observations}

Next, we summarize our observations and findings from the development and evaluation process used to analyze the novel Mojo language.

\paragraph{Observation 1: Performance portability} Without any further optimizations, Mojo's single GPU code performance is on par with AMD's HIP GPU code in all of our experiments for memory-bound kernels.
Performance differences still arise in compute-bound kernels mainly due to the lack of \texttt{fast-math} capabilities.
For the seven-point stencil operation, gaps exist between Mojo and NVIDIA's CUDA because of CUDA's additional register-level optimizations.
Surprisingly, BabelStream operations (except for \texttt{Dot}) can be further optimized by Mojo to provide a slight performance benefit versus CUDA.
This could be explained by Mojo's MLIR optimizations, which reduce the number of required register memory operations, as confirmed by our profiler analysis.
In the case of miniBUDE, there is significant sensitivity to fast-math compilation, which results in noticeable performance gaps in the Mojo implementation.
We look forward to this capability being available for these workloads in the near future.
Overall, our results suggest that Mojo can help narrow the performance gap in the Python ecosystem, but it still has gaps and a steep learning curve for low-level programming concepts.

\paragraph{Observation 2: Compile time and HPC} Mojo's compile-time nature could be a learning curve for HPC practitioners. Although HPC has been largely dominated by codes that are provided with performance-critical information at run time and do not necessarily exploit this information at compile time, Mojo might present an opportunity to explore future compiler optimizations via MLIR through AOT or JIT, as demonstrated in the literature~\cite{8945664,10.1145/3731125.3731127,9307030,10.1145/3696443.3708939}. This connection still needs to be defined, and run-time dynamic allocation is still a desired feature---even at the expense of not fully exploiting MLIR optimization. 

\paragraph{Observation 3: Python and HPC interoperability} Mojo offloads non-performance-critical code to the rich Python ecosystem.
It is still unknown how this strategy will interoperate with an HPC stack.
In particular, Mojo's interoperability with MPI (Message Passing Interface) is an open question given that Python's mpi4py is well-established~\cite{DALCIN2008655} and Python's interoperability is orthogonal to the compile-time nature of Mojo.

\paragraph{Observation 4: Libraries and ecosystem} Mojo is still a low-level programming alternative to NVIDIA's CUDA and AMD's HIP.
Many of the HPC codes rely on interactions with highly optimized high-level libraries, and support for this functionality (e.g., portable BLAS operations) is still being defined for a wider adoption of the language.
For the tools, it was extremely beneficial for our analysis, and for HPC code in general, that Mojo can generate a compiled executable that is out-of-the-box compatible with the NVIDIA \texttt{ncu} and AMD \texttt{rocprof} ecosystems without modifying any source code.

\section{Related Work}
\label{sec:Related Work}

\paragraph{Mojo} Owing to Mojo's novelty, there are very few studies of the language and its capabilities. Akre et al.~\cite{10883176} provided a recent overview of the language's potential for different areas of computing and a curated list of libraries and frameworks available on GitHub.
Their paper presents a summary of opportunities for the language and initial benchmarks that show orders-of-magnitude speedups against pure Python and Numpy code.
Raihan et al.~\cite{raihan-etal-2025-mojobench} proposed Mojobench to close the gaps for large language model--based Mojo code generation given the scant training data available for such a novel programming language.
Huang et al.~\cite{huang2025mojoframe} introduced MojoFrame, the first dataframe library to exploit performance characteristics of the language, which highlights the novel nature of Mojo's ecosystem.

In terms of high-level languages that target scientific computing, significant efforts have been made in the Python and Julia programming communities to close the gaps between high productivity and performance portability.

\paragraph{Python} Several efforts aim to enable performance-portable GPU programming without resorting to wrappers for portable code in C or C\texttt{++}.
At the vendor level, NVIDIA's recent native support for Python via \texttt{cuda-python}\footnote{\url{https://developer.nvidia.com/cuda-python}} was followed by AMD's support for interoperability, similar to HIP and CUDA, with the introduction of \texttt{hip-python-as-cuda}.\footnote{\url{https://rocm.docs.amd.com/projects/hip-python/en/latest/user_guide/2_cuda_python_interop.html}}
Mattson et al. introduced PyOMP~\cite{mattson2021pyomp}, an implementation of OpenMP for the Numba~\cite{10.1145/2833157.2833162} JIT compiler, bypassing the Python global interpreter lock. 
This work was recently extended to GPUs by using OpenMP offload~\cite{georgakoudis2024pyomp} and then benchmarked with the HeCBench~\cite{10158214} kernels. 
Similarly, Pi\~{n}eiro and Pichel proposed OMP4Py ~\cite{PINEIRO2026108035} for a pure Python implementation of OpenMP directives combining CPU multi-core with MPI parallelization.
Recently, PyOpenCL~\cite{klockner2012pycuda} and Cupy~\cite{nishino2017cupy} were expanded to provide support for AMD GPUs in addition to the well-supported NVIDIA ecosystem by passing C or C\texttt{++} custom kernel code for compilation through a strings interface. 
Similarly, pyKokkos~\cite{10.1145/3510454.3516827} provides a performance-portable interface between Python and the Kokkos library~\cite{Kokkos3} to generate C\texttt{++} Kokkos kernels and language bindings for interoperability.
More recently, Briand et al.~\cite{doi:10.1177/10943420251341179} compared the performance of Julia and several programming models, including a Kokkos.jl wrapper, against C\texttt{++} Kokkos and showed that the performance is on par with their selected computational fluid dynamics kernels.

\paragraph{Julia} The open-source LLVM-based Julia language also addresses fragmentation in the software development process, mainly targeting scientific computing and AI.
Churavy et al.~\cite{JuliaHPC} summarized efforts to establish a unifying ecosystem for large-scale computing.
Valero-Lara et al.~\cite{10820713} introduced JACC, a recently developed high-level, performance-portable library that abstracts away low-level, hardware-specific options on CPUs/GPUs while exhibiting nearly zero overhead versus Julia's vendor-specific CUDA.jl~\cite{8471188}, AMDGPU.jl~\cite{julian_samaroo_2023_10040461}, and oneAPI.jl~\cite{besard_2025_14931098} back ends. 
Similarly, Churavy et al. developed KernelAbstractions.jl~\cite{churavy_2025_14724294}, which enables portable GPU kernel programming by using a fine-granularity parallelization approach similar to the CUDA and HIP models.
Lin and McIntosh-Smith~\cite{9652798} analyzed the performance of Julia's programming models for BabelStream and miniBUDE on CPUs and GPUs.
Their study showed that Julia is on par or slightly behind compiled languages on several CPUs and GPUs.
Godoy et al.~\cite{godoy2023evaluating} evaluated the performance portability of Julia, Numba, and Kokkos programming models across CPUs and NVIDIA and AMD GPUs.
The work demonstrated that Numba was lacking in performance and portability on AMD GPUs, but Julia's GPU capabilities showed significant potential---even with some gaps in performance for a simple matrix multiplication kernel.
De la Calle et al.~\cite{DELACALLE2025107813} introduced the Juliana tool to add GPU performance portability in Julia by porting CUDA.jl code to KernelAbstractions.jl, which introduces a small overhead on several kernels, including miniBUDE and BabelStream.

\paragraph{Other efforts} Efforts to narrow gaps of productivity, performance, and portability have long existed in HPC.
The Defense Advanced Research Projects Agency's High Productivity Computing Systems~\cite{DONGARRA20081} program funded three languages---Chapel~\cite{chapel}, IBM's X10~\cite{x10}, and Sun's Fortress~\cite{fortress}---and favored a partitioned global address space model.
Only Chapel remains in active development as it evolves toward vendor-neutral GPU programming~\cite{kayraklioglu2024productive}.
On the proprietary side, parallel Matlab~\cite{1386655} is an early effort to leverage the popular Matlab language and parallel computing.

Hence, Mojo aligns with the search for productive, vendor-neutral programming models to drive future extremely heterogeneous computing systems~\cite{osti_1473756}.

\section{Conclusions}
\label{sec:Conclusions}

We present our early experiences with the novel Mojo programming language. In particular, we describe the added capability to write performance-portable kernels for NVIDIA and AMD GPUs.
We targeted four common science kernels---(i) seven-point stencil (memory-bound), (ii) BabelStream (memory-bound), (iii) Hartree--Fock (compute-bound + atomic), and (iv) miniBUDE (compute-bound)---and compared the same Mojo code with CUDA and HIP on modern NVIDIA H100 and AMD MI300A GPUs, respectively.

Our results indicate that the performance of the Mojo low-level implementation is on par with AMD's HIP for memory--bandwidth bound cases. 
We observed lower performance when comparing Mojo against CUDA-specific seven-point stencil (89\%) and the shared-memory operations in the \texttt{Dot} kernel (78\%) in BabelStream.
On the plus side, Mojo is highly optimized for simple 1D array operations in BabelStream and even slightly surpassed CUDA's performance thanks to fewer cached memory operations. 
The differences between Mojo and the CUDA implementation were observed by comparing profiling information from vendor tools.
These differences need to be understood at lower compiler levels to avoid sacrificing the potential of Mojo's portability.
For the compute-bound workloads in miniBUDE and Hartree--Fock, which included atomic operations, Mojo still exhibits several performance gaps versus CUDA and HIP.
The timeline for adding the fundamental \texttt{fast-math} optimizations to Mojo is still an unknown, as these optimizations are critical to the tested compute-bound kernels.
Additionally, atomic operations must be further analyzed and better understood given Mojo's peculiar results of overperforming CUDA and underperforming HIP in these tasks.

Based on the performance results, we applied a performance-portability metric, but we expect the actual performance portability to evolve as the language matures, adds features to fully leverage MLIR, and integrates feedback from the open-source HPC community.

Unlike other programming models that attempt to speed up Python or compile to LLVM from the ground up (e.g., Julia, Rust), Mojo closes performance gaps by offering a superset language that combines rich compile-time metaprogramming via MLIR, performance portable GPU programming capabilities in their standard library, and run-time interoperability with the Python ecosystem.
Our expectation is that Mojo will continue to evolve under this programming paradigm---with the promise of becoming open-source by 2026\footnote{\url{https://docs.modular.com/mojo/faq}}---as it moves toward becoming a viable HPC~+~AI programming language.

\begin{acks}
This work was supported in part by the U.S. Department of Energy, Office of Science, Office of Advanced Scientific Computing Research’s Computer Science Competitive Portfolios program, MAGMA/Fairbanks project; and the Next Generation of Scientific Software Technologies program, PESO and S4PST projects.
This research used resources of the Oak Ridge Leadership Computing Facility and the Experimental Computing Laboratory at the Oak Ridge National Laboratory, which are supported by the Office of Science of the US Department of Energy under Contract No. DE-AC05-00OR22725. This work was supported in part by the U.S. Department of Energy, Office of Science, Office of Workforce Development for Teachers and Scientists (WDTS) under the Science Undergraduate Laboratory Internships Program (SULI).
WFG and TM would like to thank Alex Smith from the University of Wisconsin-Madison for providing the CUDA and HIP ports of Hartree-Fock.
\end{acks}

\bibliographystyle{ACM-Reference-Format}
\bibliography{references}


\begin{thebibliography}{58}


\ifx \showCODEN    \undefined \def \showCODEN     #1{\unskip}     \fi
\ifx \showDOI      \undefined \def \showDOI       #1{#1}\fi
\ifx \showISBNx    \undefined \def \showISBNx     #1{\unskip}     \fi
\ifx \showISBNxiii \undefined \def \showISBNxiii  #1{\unskip}     \fi
\ifx \showISSN     \undefined \def \showISSN      #1{\unskip}     \fi
\ifx \showLCCN     \undefined \def \showLCCN      #1{\unskip}     \fi
\ifx \shownote     \undefined \def \shownote      #1{#1}          \fi
\ifx \showarticletitle \undefined \def \showarticletitle #1{#1}   \fi
\ifx \showURL      \undefined \def \showURL       {\relax}        \fi
\providecommand\bibfield[2]{#2}
\providecommand\bibinfo[2]{#2}
\providecommand\natexlab[1]{#1}
\providecommand\showeprint[2][]{arXiv:#2}

\bibitem[Akre and Pacharaney(2025)]%
        {10883176}
\bibfield{author}{\bibinfo{person}{Parth~Dhananjay Akre} {and} \bibinfo{person}{Utkarsha Pacharaney}.} \bibinfo{year}{2025}\natexlab{}.
\newblock \showarticletitle{A Comprehensive Review of Mojo: A High-Performance Programming Language}. In \bibinfo{booktitle}{\emph{2025 6th International Conference on Mobile Computing and Sustainable Informatics (ICMCSI)}}. \bibinfo{pages}{861--867}.
\newblock
\urldef\tempurl%
\url{https://doi.org/10.1109/ICMCSI64620.2025.10883176}
\showDOI{\tempurl}


\bibitem[Allen et~al\mbox{.}(2005)]%
        {fortress}
\bibfield{author}{\bibinfo{person}{Eric Allen}, \bibinfo{person}{David Chase}, \bibinfo{person}{Joe Hallett}, \bibinfo{person}{Victor Luchangco}, \bibinfo{person}{Jan-Willem Maessen}, \bibinfo{person}{Sukyoung Ryu}, \bibinfo{person}{Guy~L Steele~Jr}, \bibinfo{person}{Sam Tobin-Hochstadt}, \bibinfo{person}{Joao Dias}, \bibinfo{person}{Carl Eastlund}, {et~al\mbox{.}}} \bibinfo{year}{2005}\natexlab{}.
\newblock \showarticletitle{The {F}ortress language specification}.
\newblock \bibinfo{journal}{\emph{Sun Microsystems}} \bibinfo{volume}{139}, \bibinfo{number}{140} (\bibinfo{year}{2005}), \bibinfo{pages}{116}.
\newblock


\bibitem[Alpay and Heuveline(2025)]%
        {10.1145/3731125.3731127}
\bibfield{author}{\bibinfo{person}{Aksel Alpay} {and} \bibinfo{person}{Vincent Heuveline}.} \bibinfo{year}{2025}\natexlab{}.
\newblock \showarticletitle{Adaptivity in AdaptiveCpp: Optimizing Performance by Leveraging Runtime Information During JIT-Compilation}. In \bibinfo{booktitle}{\emph{Proceedings of the 13th International Workshop on OpenCL and SYCL}} \emph{(\bibinfo{series}{IWOCL '25})}. \bibinfo{publisher}{Association for Computing Machinery}, \bibinfo{address}{New York, NY, USA}, Article \bibinfo{articleno}{2}, \bibinfo{numpages}{12}~pages.
\newblock
\showISBNx{9798400713606}
\urldef\tempurl%
\url{https://doi.org/10.1145/3731125.3731127}
\showDOI{\tempurl}


\bibitem[Awar et~al\mbox{.}(2022)]%
        {10.1145/3510454.3516827}
\bibfield{author}{\bibinfo{person}{Nader~Al Awar}, \bibinfo{person}{Neil Mehta}, \bibinfo{person}{Steven Zhu}, \bibinfo{person}{George Biros}, {and} \bibinfo{person}{Milos Gligoric}.} \bibinfo{year}{2022}\natexlab{}.
\newblock \showarticletitle{PyKokkos: performance portable kernels in Python}. In \bibinfo{booktitle}{\emph{Proceedings of the ACM/IEEE 44th International Conference on Software Engineering: Companion Proceedings}} (Pittsburgh, Pennsylvania) \emph{(\bibinfo{series}{ICSE '22})}. \bibinfo{publisher}{Association for Computing Machinery}, \bibinfo{address}{New York, NY, USA}, \bibinfo{pages}{164–167}.
\newblock
\showISBNx{9781450392235}
\urldef\tempurl%
\url{https://doi.org/10.1145/3510454.3516827}
\showDOI{\tempurl}


\bibitem[Beckingsale et~al\mbox{.}(2019)]%
        {Raja}
\bibfield{author}{\bibinfo{person}{David~A. Beckingsale} {et~al\mbox{.}}} \bibinfo{year}{2019}\natexlab{}.
\newblock \showarticletitle{RAJA: Portable Performance for Large-Scale Scientific Applications}. In \bibinfo{booktitle}{\emph{IEEE/ACM International Workshop on Performance, Portability and Productivity in HPC (P3HPC)}}. \bibinfo{pages}{71--81}.
\newblock
\urldef\tempurl%
\url{https://doi.org/10.1109/P3HPC49587.2019.00012}
\showDOI{\tempurl}


\bibitem[Besard(2025)]%
        {besard_2025_14931098}
\bibfield{author}{\bibinfo{person}{Tim Besard}.} \bibinfo{year}{2025}\natexlab{}.
\newblock \bibinfo{booktitle}{\emph{oneAPI.jl}}.
\newblock
\urldef\tempurl%
\url{https://doi.org/10.5281/zenodo.14931098}
\showDOI{\tempurl}


\bibitem[Besard et~al\mbox{.}(2019)]%
        {8471188}
\bibfield{author}{\bibinfo{person}{Tim Besard}, \bibinfo{person}{Christophe Foket}, {and} \bibinfo{person}{Bjorn De~Sutter}.} \bibinfo{year}{2019}\natexlab{}.
\newblock \showarticletitle{Effective Extensible Programming: Unleashing Julia on GPUs}.
\newblock \bibinfo{journal}{\emph{IEEE Transactions on Parallel and Distributed Systems}} \bibinfo{volume}{30}, \bibinfo{number}{4} (\bibinfo{year}{2019}), \bibinfo{pages}{827--841}.
\newblock
\urldef\tempurl%
\url{https://doi.org/10.1109/TPDS.2018.2872064}
\showDOI{\tempurl}


\bibitem[Bezanson et~al\mbox{.}(2017)]%
        {Bezanson2017-ca}
\bibfield{author}{\bibinfo{person}{Jeff Bezanson} {et~al\mbox{.}}} \bibinfo{year}{2017}\natexlab{}.
\newblock \showarticletitle{Julia: A Fresh Approach to Numerical Computing}.
\newblock \bibinfo{journal}{\emph{SIAM Rev.}} \bibinfo{volume}{59}, \bibinfo{number}{1} (\bibinfo{date}{Jan.} \bibinfo{year}{2017}), \bibinfo{pages}{65--98}.
\newblock
\showISSN{0036-1445}
\urldef\tempurl%
\url{https://doi.org/10.1137/141000671}
\showDOI{\tempurl}


\bibitem[Briand et~al\mbox{.}(2025)]%
        {doi:10.1177/10943420251341179}
\bibfield{author}{\bibinfo{person}{Luc Briand}, \bibinfo{person}{Hervé Jourdren}, {and} \bibinfo{person}{Marc Pérache}.} \bibinfo{year}{2025}\natexlab{}.
\newblock \showarticletitle{Julia versus C++ Kokkos for performance portable Cartesian CFD solvers on heterogeneous architectures}.
\newblock \bibinfo{journal}{\emph{The International Journal of High Performance Computing Applications}} \bibinfo{volume}{39}, \bibinfo{number}{4} (\bibinfo{year}{2025}), \bibinfo{pages}{481--501}.
\newblock
\urldef\tempurl%
\url{https://doi.org/10.1177/10943420251341179}
\showDOI{\tempurl}


\bibitem[Chamberlain et~al\mbox{.}(2007)]%
        {chapel}
\bibfield{author}{\bibinfo{person}{B.L. Chamberlain}, \bibinfo{person}{D. Callahan}, {and} \bibinfo{person}{H.P. Zima}.} \bibinfo{year}{2007}\natexlab{}.
\newblock \showarticletitle{Parallel {P}rogrammability and the {C}hapel {L}anguage}.
\newblock \bibinfo{journal}{\emph{The International Journal of High Performance Computing Applications}} \bibinfo{volume}{21}, \bibinfo{number}{3} (\bibinfo{year}{2007}), \bibinfo{pages}{291--312}.
\newblock
\urldef\tempurl%
\url{https://doi.org/10.1177/1094342007078442}
\showDOI{\tempurl}


\bibitem[Choy and Edelman(2005)]%
        {1386655}
\bibfield{author}{\bibinfo{person}{R. Choy} {and} \bibinfo{person}{A. Edelman}.} \bibinfo{year}{2005}\natexlab{}.
\newblock \showarticletitle{Parallel MATLAB: Doing it Right}.
\newblock \bibinfo{journal}{\emph{Proc. IEEE}} \bibinfo{volume}{93}, \bibinfo{number}{2} (\bibinfo{year}{2005}), \bibinfo{pages}{331--341}.
\newblock
\urldef\tempurl%
\url{https://doi.org/10.1109/JPROC.2004.840490}
\showDOI{\tempurl}


\bibitem[Churavy(2025)]%
        {churavy_2025_14724294}
\bibfield{author}{\bibinfo{person}{Valentin Churavy}.} \bibinfo{year}{2025}\natexlab{}.
\newblock \bibinfo{booktitle}{\emph{KernelAbstractions.jl}}.
\newblock
\urldef\tempurl%
\url{https://doi.org/10.5281/zenodo.14724294}
\showDOI{\tempurl}


\bibitem[Churavy et~al\mbox{.}(2022)]%
        {JuliaHPC}
\bibfield{author}{\bibinfo{person}{Valentin Churavy} {et~al\mbox{.}}} \bibinfo{year}{2022}\natexlab{}.
\newblock \bibinfo{title}{Bridging HPC Communities through the Julia Programming Language}.
\newblock
\newblock
\showeprint[arxiv]{2211.02740}~[cs.DC]


\bibitem[Dalcín et~al\mbox{.}(2008)]%
        {DALCIN2008655}
\bibfield{author}{\bibinfo{person}{Lisandro Dalcín}, \bibinfo{person}{Rodrigo Paz}, \bibinfo{person}{Mario Storti}, {and} \bibinfo{person}{Jorge D’Elía}.} \bibinfo{year}{2008}\natexlab{}.
\newblock \showarticletitle{MPI for Python: Performance improvements and MPI-2 extensions}.
\newblock \bibinfo{journal}{\emph{J. Parallel and Distrib. Comput.}} \bibinfo{volume}{68}, \bibinfo{number}{5} (\bibinfo{year}{2008}), \bibinfo{pages}{655--662}.
\newblock
\showISSN{0743-7315}
\urldef\tempurl%
\url{https://doi.org/10.1016/j.jpdc.2007.09.005}
\showDOI{\tempurl}


\bibitem[Datta et~al\mbox{.}(2008)]%
        {5222004}
\bibfield{author}{\bibinfo{person}{Kaushik Datta}, \bibinfo{person}{Mark Murphy}, \bibinfo{person}{Vasily Volkov}, \bibinfo{person}{Samuel Williams}, \bibinfo{person}{Jonathan Carter}, \bibinfo{person}{Leonid Oliker}, \bibinfo{person}{David Patterson}, \bibinfo{person}{John Shalf}, {and} \bibinfo{person}{Katherine Yelick}.} \bibinfo{year}{2008}\natexlab{}.
\newblock \showarticletitle{Stencil computation optimization and auto-tuning on state-of-the-art multicore architectures}. In \bibinfo{booktitle}{\emph{SC '08: Proceedings of the 2008 ACM/IEEE Conference on Supercomputing}}. \bibinfo{pages}{1--12}.
\newblock
\urldef\tempurl%
\url{https://doi.org/10.1109/SC.2008.5222004}
\showDOI{\tempurl}


\bibitem[{de la Calle} and García(2025)]%
        {DELACALLE2025107813}
\bibfield{author}{\bibinfo{person}{Enrique {de la Calle}} {and} \bibinfo{person}{Carlos García}.} \bibinfo{year}{2025}\natexlab{}.
\newblock \showarticletitle{Evaluation of Juliana Tool: A translator for Julia’s CUDA.jl code into KernelAbstraction.jl}.
\newblock \bibinfo{journal}{\emph{Future Generation Computer Systems}} (\bibinfo{year}{2025}), \bibinfo{pages}{107813}.
\newblock
\showISSN{0167-739X}
\urldef\tempurl%
\url{https://doi.org/10.1016/j.future.2025.107813}
\showDOI{\tempurl}


\bibitem[Deakin et~al\mbox{.}(2018)]%
        {deakin2018evaluating}
\bibfield{author}{\bibinfo{person}{Tom Deakin} {et~al\mbox{.}}} \bibinfo{year}{2018}\natexlab{}.
\newblock \showarticletitle{Evaluating attainable memory bandwidth of parallel programming models via BabelStream}.
\newblock \bibinfo{journal}{\emph{International Journal of Computational Science and Engineering}} \bibinfo{volume}{17}, \bibinfo{number}{3} (\bibinfo{year}{2018}), \bibinfo{pages}{247--262}.
\newblock


\bibitem[Deakin and McIntosh-Smith(2020)]%
        {deakin2020evaluating}
\bibfield{author}{\bibinfo{person}{Tom Deakin} {and} \bibinfo{person}{Simon McIntosh-Smith}.} \bibinfo{year}{2020}\natexlab{}.
\newblock \showarticletitle{Evaluating the performance of HPC-style SYCL applications}. In \bibinfo{booktitle}{\emph{International Workshop on OpenCL}}. \bibinfo{pages}{1--11}.
\newblock


\bibitem[Dongarra et~al\mbox{.}(2008)]%
        {DONGARRA20081}
\bibfield{author}{\bibinfo{person}{Jack Dongarra}, \bibinfo{person}{Robert Graybill}, \bibinfo{person}{William Harrod}, \bibinfo{person}{Robert Lucas}, \bibinfo{person}{Ewing Lusk}, \bibinfo{person}{Piotr Luszczek}, \bibinfo{person}{Janice Mcmahon}, \bibinfo{person}{Allan Snavely}, \bibinfo{person}{Jeffrey Vetter}, \bibinfo{person}{Katherine Yelick}, \bibinfo{person}{Sadaf Alam}, \bibinfo{person}{Roy Campbell}, \bibinfo{person}{Laura Carrington}, \bibinfo{person}{Tzu-Yi Chen}, \bibinfo{person}{Omid Khalili}, \bibinfo{person}{Jeremy Meredith}, {and} \bibinfo{person}{Mustafa Tikir}.} \bibinfo{year}{2008}\natexlab{}.
\newblock \showarticletitle{DARPA's HPCS Program: History, Models, Tools, Languages}.
\newblock In \bibinfo{booktitle}{\emph{Advances in COMPUTERS}}. \bibinfo{series}{Advances in Computers}, Vol.~\bibinfo{volume}{72}. \bibinfo{publisher}{Elsevier}, \bibinfo{pages}{1--100}.
\newblock
\showISSN{0065-2458}
\urldef\tempurl%
\url{https://doi.org/10.1016/S0065-2458(08)00001-6}
\showDOI{\tempurl}


\bibitem[Dongarra et~al\mbox{.}(2003)]%
        {dongarra2003linpack}
\bibfield{author}{\bibinfo{person}{Jack~J Dongarra}, \bibinfo{person}{Piotr Luszczek}, {and} \bibinfo{person}{Antoine Petitet}.} \bibinfo{year}{2003}\natexlab{}.
\newblock \showarticletitle{The LINPACK benchmark: past, present and future}.
\newblock \bibinfo{journal}{\emph{Concurrency and Computation: practice and experience}} \bibinfo{volume}{15}, \bibinfo{number}{9} (\bibinfo{year}{2003}), \bibinfo{pages}{803--820}.
\newblock


\bibitem[Engelke and Schulz(2020)]%
        {9307030}
\bibfield{author}{\bibinfo{person}{Alexis Engelke} {and} \bibinfo{person}{Martin Schulz}.} \bibinfo{year}{2020}\natexlab{}.
\newblock \showarticletitle{Robust Practical Binary Optimization at Run-time using LLVM}. In \bibinfo{booktitle}{\emph{2020 IEEE/ACM 6th Workshop on the LLVM Compiler Infrastructure in HPC (LLVM-HPC) and Workshop on Hierarchical Parallelism for Exascale Computing (HiPar)}}. \bibinfo{pages}{56--64}.
\newblock
\urldef\tempurl%
\url{https://doi.org/10.1109/LLVMHPCHiPar51896.2020.00011}
\showDOI{\tempurl}


\bibitem[Finkel et~al\mbox{.}(2019)]%
        {8945664}
\bibfield{author}{\bibinfo{person}{Hal Finkel}, \bibinfo{person}{David Poliakoff}, \bibinfo{person}{Jean-Sylvain Camier}, {and} \bibinfo{person}{David~F. Richards}.} \bibinfo{year}{2019}\natexlab{}.
\newblock \showarticletitle{ClangJIT: Enhancing C++ with Just-in-Time Compilation}. In \bibinfo{booktitle}{\emph{2019 IEEE/ACM International Workshop on Performance, Portability and Productivity in HPC (P3HPC)}}. \bibinfo{pages}{82--95}.
\newblock
\urldef\tempurl%
\url{https://doi.org/10.1109/P3HPC49587.2019.00013}
\showDOI{\tempurl}


\bibitem[Fletcher et~al\mbox{.}(2021)]%
        {osti_1783174}
\bibfield{author}{\bibinfo{person}{Graham~D Fletcher} {et~al\mbox{.}}} \bibinfo{year}{2021}\natexlab{}.
\newblock \bibinfo{title}{{Basic Hartree-Fock Proxy Application}}.
\newblock
\newblock
\urldef\tempurl%
\url{https://doi.org/10.11578/dc.20210517.2}
\showDOI{\tempurl}


\bibitem[Georgakoudis et~al\mbox{.}(2024)]%
        {georgakoudis2024pyomp}
\bibfield{author}{\bibinfo{person}{Giorgis Georgakoudis}, \bibinfo{person}{Todd~A Anderson}, \bibinfo{person}{Stuart Archibald}, \bibinfo{person}{Bronis de Supinski}, {and} \bibinfo{person}{Timothy~G Mattson}.} \bibinfo{year}{2024}\natexlab{}.
\newblock \bibinfo{booktitle}{\emph{PyOMP: Programming GPUs with OpenMP and Python}}.
\newblock \bibinfo{type}{{T}echnical {R}eport}. \bibinfo{institution}{Lawrence Livermore National Laboratory (LLNL), Livermore, CA (United States)}.
\newblock


\bibitem[Georgakoudis et~al\mbox{.}(2025)]%
        {10.1145/3696443.3708939}
\bibfield{author}{\bibinfo{person}{Giorgis Georgakoudis}, \bibinfo{person}{Konstantinos Parasyris}, {and} \bibinfo{person}{David Beckingsale}.} \bibinfo{year}{2025}\natexlab{}.
\newblock \showarticletitle{Proteus: Portable Runtime Optimization of GPU Kernel Execution with Just-in-Time Compilation}. In \bibinfo{booktitle}{\emph{Proceedings of the 23rd ACM/IEEE International Symposium on Code Generation and Optimization}} (Las Vegas, NV, USA) \emph{(\bibinfo{series}{CGO '25})}. \bibinfo{publisher}{Association for Computing Machinery}, \bibinfo{address}{New York, NY, USA}, \bibinfo{pages}{507–522}.
\newblock
\showISBNx{9798400712753}
\urldef\tempurl%
\url{https://doi.org/10.1145/3696443.3708939}
\showDOI{\tempurl}


\bibitem[Godoy and Liu(2011)]%
        {doi:10.1080/10407790.2010.541359}
\bibfield{author}{\bibinfo{person}{William~F. Godoy} {and} \bibinfo{person}{Xu Liu}.} \bibinfo{year}{2011}\natexlab{}.
\newblock \showarticletitle{Introduction of Parallel GPGPU Acceleration Algorithms for the Solution of Radiative Transfer}.
\newblock \bibinfo{journal}{\emph{Numerical Heat Transfer, Part B: Fundamentals}} \bibinfo{volume}{59}, \bibinfo{number}{1} (\bibinfo{year}{2011}), \bibinfo{pages}{1--25}.
\newblock
\urldef\tempurl%
\url{https://doi.org/10.1080/10407790.2010.541359}
\showDOI{\tempurl}


\bibitem[Godoy et~al\mbox{.}(2023)]%
        {godoy2023evaluating}
\bibfield{author}{\bibinfo{person}{William~F Godoy}, \bibinfo{person}{Pedro Valero-Lara}, \bibinfo{person}{T~Elise Dettling}, \bibinfo{person}{Christian Trefftz}, \bibinfo{person}{Ian Jorquera}, \bibinfo{person}{Thomas Sheehy}, \bibinfo{person}{Ross~G Miller}, \bibinfo{person}{Marc Gonzalez-Tallada}, \bibinfo{person}{Jeffrey~S Vetter}, {and} \bibinfo{person}{Valentin Churavy}.} \bibinfo{year}{2023}\natexlab{}.
\newblock \showarticletitle{Evaluating performance and portability of high-level programming models: Julia, Python/Numba, and Kokkos on exascale nodes}. In \bibinfo{booktitle}{\emph{2023 IEEE international parallel and distributed processing symposium workshops (IPDPSW)}}. IEEE, \bibinfo{pages}{373--382}.
\newblock
\urldef\tempurl%
\url{https://doi.org/10.1109/IPDPSW59300.2023.00068}
\showDOI{\tempurl}


\bibitem[Holewinski et~al\mbox{.}(2012)]%
        {10.1145/2304576.2304619}
\bibfield{author}{\bibinfo{person}{Justin Holewinski}, \bibinfo{person}{Louis-No\"{e}l Pouchet}, {and} \bibinfo{person}{P. Sadayappan}.} \bibinfo{year}{2012}\natexlab{}.
\newblock \showarticletitle{High-Performance Code Generation for Stencil Computations on GPU Architectures}. In \bibinfo{booktitle}{\emph{Proceedings of the 26th ACM International Conference on Supercomputing}} (San Servolo Island, Venice, Italy) \emph{(\bibinfo{series}{ICS '12})}. \bibinfo{publisher}{Association for Computing Machinery}, \bibinfo{address}{New York, NY, USA}, \bibinfo{pages}{311–320}.
\newblock
\showISBNx{9781450313162}
\urldef\tempurl%
\url{https://doi.org/10.1145/2304576.2304619}
\showDOI{\tempurl}


\bibitem[Huang et~al\mbox{.}(2025)]%
        {huang2025mojoframe}
\bibfield{author}{\bibinfo{person}{Shengya Huang}, \bibinfo{person}{Zhaoheng Li}, \bibinfo{person}{Derek Werner}, {and} \bibinfo{person}{Yongjoo Park}.} \bibinfo{year}{2025}\natexlab{}.
\newblock \showarticletitle{MojoFrame: Dataframe Library in Mojo Language}.
\newblock \bibinfo{journal}{\emph{arXiv preprint arXiv:2505.04080}} (\bibinfo{year}{2025}).
\newblock


\bibitem[Jin and Vetter(2023)]%
        {10158214}
\bibfield{author}{\bibinfo{person}{Zheming Jin} {and} \bibinfo{person}{Jeffrey~S. Vetter}.} \bibinfo{year}{2023}\natexlab{}.
\newblock \showarticletitle{A Benchmark Suite for Improving Performance Portability of the SYCL Programming Model}. In \bibinfo{booktitle}{\emph{2023 IEEE International Symposium on Performance Analysis of Systems and Software (ISPASS)}}. \bibinfo{pages}{325--327}.
\newblock
\urldef\tempurl%
\url{https://doi.org/10.1109/ISPASS57527.2023.00041}
\showDOI{\tempurl}


\bibitem[Kayraklioglu and Stone(2024)]%
        {kayraklioglu2024productive}
\bibfield{author}{\bibinfo{person}{Engin Kayraklioglu} {and} \bibinfo{person}{Andy Stone}.} \bibinfo{year}{2024}\natexlab{}.
\newblock \showarticletitle{Productive, Vendor-Neutral GPU Programming Using Chapel}. In \bibinfo{booktitle}{\emph{SC24-W: Workshops of the International Conference for High Performance Computing, Networking, Storage and Analysis}}. IEEE, \bibinfo{pages}{1914--1922}.
\newblock


\bibitem[Kl{\"o}ckner et~al\mbox{.}(2012)]%
        {klockner2012pycuda}
\bibfield{author}{\bibinfo{person}{Andreas Kl{\"o}ckner}, \bibinfo{person}{Nicolas Pinto}, \bibinfo{person}{Yunsup Lee}, \bibinfo{person}{Bryan Catanzaro}, \bibinfo{person}{Paul Ivanov}, {and} \bibinfo{person}{Ahmed Fasih}.} \bibinfo{year}{2012}\natexlab{}.
\newblock \showarticletitle{PyCUDA and PyOpenCL: A scripting-based approach to GPU run-time code generation}.
\newblock \bibinfo{journal}{\emph{Parallel computing}} \bibinfo{volume}{38}, \bibinfo{number}{3} (\bibinfo{year}{2012}), \bibinfo{pages}{157--174}.
\newblock


\bibitem[Krotkiewski and Dabrowski(2013)]%
        {KROTKIEWSKI2013533}
\bibfield{author}{\bibinfo{person}{Marcin Krotkiewski} {and} \bibinfo{person}{Marcin Dabrowski}.} \bibinfo{year}{2013}\natexlab{}.
\newblock \showarticletitle{Efficient 3D stencil computations using CUDA}.
\newblock \bibinfo{journal}{\emph{Parallel Comput.}} \bibinfo{volume}{39}, \bibinfo{number}{10} (\bibinfo{year}{2013}), \bibinfo{pages}{533--548}.
\newblock
\showISSN{0167-8191}
\urldef\tempurl%
\url{https://doi.org/10.1016/j.parco.2013.08.002}
\showDOI{\tempurl}


\bibitem[Lam et~al\mbox{.}(2015)]%
        {10.1145/2833157.2833162}
\bibfield{author}{\bibinfo{person}{Siu~Kwan Lam}, \bibinfo{person}{Antoine Pitrou}, {and} \bibinfo{person}{Stanley Seibert}.} \bibinfo{year}{2015}\natexlab{}.
\newblock \showarticletitle{Numba: a LLVM-based Python JIT compiler}. In \bibinfo{booktitle}{\emph{Proceedings of the Second Workshop on the LLVM Compiler Infrastructure in HPC}} (Austin, Texas) \emph{(\bibinfo{series}{LLVM '15})}. \bibinfo{publisher}{Association for Computing Machinery}, \bibinfo{address}{New York, NY, USA}, Article \bibinfo{articleno}{7}, \bibinfo{numpages}{6}~pages.
\newblock
\showISBNx{9781450340052}
\urldef\tempurl%
\url{https://doi.org/10.1145/2833157.2833162}
\showDOI{\tempurl}


\bibitem[Lattner and Adve(2004)]%
        {lattner2004llvm}
\bibfield{author}{\bibinfo{person}{Chris Lattner} {and} \bibinfo{person}{Vikram Adve}.} \bibinfo{year}{2004}\natexlab{}.
\newblock \showarticletitle{{LLVM: A compilation framework for lifelong program analysis \& transformation}}. In \bibinfo{booktitle}{\emph{International Symposium on Code Generation and Optimization, 2004. CGO 2004.}} IEEE, \bibinfo{pages}{75--86}.
\newblock


\bibitem[Lattner and Pienaar(2019)]%
        {lattner2019mlir}
\bibfield{author}{\bibinfo{person}{Chris Lattner} {and} \bibinfo{person}{Jacques Pienaar}.} \bibinfo{year}{2019}\natexlab{}.
\newblock \showarticletitle{{MLIR primer: A compiler infrastructure for the end of Moore’s law}}. In \bibinfo{booktitle}{\emph{Compilers for Machine Learning. C4ML workshop at CGO 2019}}. \bibinfo{pages}{100}.
\newblock


\bibitem[Lin and McIntosh-Smith(2021)]%
        {9652798}
\bibfield{author}{\bibinfo{person}{Wei-Chen Lin} {and} \bibinfo{person}{Simon McIntosh-Smith}.} \bibinfo{year}{2021}\natexlab{}.
\newblock \showarticletitle{{Comparing Julia to Performance Portable Parallel Programming Models for HPC}}. In \bibinfo{booktitle}{\emph{International Workshop on Performance Modeling, Benchmarking and Simulation of High Performance Computer Systems (PMBS)}}. \bibinfo{pages}{94--105}.
\newblock
\urldef\tempurl%
\url{https://doi.org/10.1109/PMBS54543.2021.00016}
\showDOI{\tempurl}


\bibitem[Marowka(2021)]%
        {9407113}
\bibfield{author}{\bibinfo{person}{Ami Marowka}.} \bibinfo{year}{2021}\natexlab{}.
\newblock \showarticletitle{Toward a Better Performance Portability Metric}. In \bibinfo{booktitle}{\emph{2021 29th Euromicro International Conference on Parallel, Distributed and Network-Based Processing (PDP)}}. \bibinfo{pages}{181--184}.
\newblock
\urldef\tempurl%
\url{https://doi.org/10.1109/PDP52278.2021.00036}
\showDOI{\tempurl}


\bibitem[Marowka(2025)]%
        {MAROWKA2025107826}
\bibfield{author}{\bibinfo{person}{Ami Marowka}.} \bibinfo{year}{2025}\natexlab{}.
\newblock \showarticletitle{Portability efficiency approach for calculating performance portability}.
\newblock \bibinfo{journal}{\emph{Future Generation Computer Systems}}  \bibinfo{volume}{170} (\bibinfo{year}{2025}), \bibinfo{pages}{107826}.
\newblock
\showISSN{0167-739X}
\urldef\tempurl%
\url{https://doi.org/10.1016/j.future.2025.107826}
\showDOI{\tempurl}


\bibitem[Matsakis and Klock(2014)]%
        {matsakis2014rust}
\bibfield{author}{\bibinfo{person}{Nicholas~D Matsakis} {and} \bibinfo{person}{Felix~S Klock}.} \bibinfo{year}{2014}\natexlab{}.
\newblock \showarticletitle{{The Rust Language}}. In \bibinfo{booktitle}{\emph{Proceedings of the 2014 ACM SIGAda annual conference on High integrity language technology}}. \bibinfo{pages}{103--104}.
\newblock


\bibitem[Mattson et~al\mbox{.}(2021)]%
        {mattson2021pyomp}
\bibfield{author}{\bibinfo{person}{Timothy~G Mattson}, \bibinfo{person}{Todd~A Anderson}, {and} \bibinfo{person}{Giorgis Georgakoudis}.} \bibinfo{year}{2021}\natexlab{}.
\newblock \showarticletitle{Pyomp: Multithreaded parallel programming in python}.
\newblock \bibinfo{journal}{\emph{Computing in Science \& Engineering}} \bibinfo{volume}{23}, \bibinfo{number}{6} (\bibinfo{year}{2021}), \bibinfo{pages}{77--80}.
\newblock


\bibitem[Modular({[n.\,d.]})]%
        {Mojo}
\bibfield{author}{\bibinfo{person}{Modular}.} \bibinfo{year}{[n.\,d.]}\natexlab{}.
\newblock \bibinfo{title}{Mojo, Powerful CPU+GPU Programming}.
\newblock
\newblock
\urldef\tempurl%
\url{https://www.modular.com/mojo}
\showURL{%
\tempurl}


\bibitem[Nishino and Loomis(2017)]%
        {nishino2017cupy}
\bibfield{author}{\bibinfo{person}{ROYUD Nishino} {and} \bibinfo{person}{Shohei Hido~Crissman Loomis}.} \bibinfo{year}{2017}\natexlab{}.
\newblock \showarticletitle{Cupy: A numpy-compatible library for nvidia gpu calculations}.
\newblock \bibinfo{journal}{\emph{31st confernce on neural information processing systems}} \bibinfo{volume}{151}, \bibinfo{number}{7} (\bibinfo{year}{2017}).
\newblock


\bibitem[{OpenMP Architecture Review Board}(2021)]%
        {openmp}
\bibfield{author}{\bibinfo{person}{{OpenMP Architecture Review Board}}.} \bibinfo{year}{2021}\natexlab{}.
\newblock \bibinfo{title}{{OpenMP} Application Program Interface Version 5.2}.
\newblock
\newblock
\urldef\tempurl%
\url{https://www.openmp.org/wp-content/uploads/OpenMP-API-Specification-5-2.pdf}
\showURL{%
\tempurl}


\bibitem[Pennycook and Sewall(2021)]%
        {pennycook2021revisiting}
\bibfield{author}{\bibinfo{person}{S~John Pennycook} {and} \bibinfo{person}{Jason~D Sewall}.} \bibinfo{year}{2021}\natexlab{}.
\newblock \showarticletitle{Revisiting a metric for performance portability}. In \bibinfo{booktitle}{\emph{2021 International Workshop on Performance, Portability and Productivity in HPC (P3HPC)}}. IEEE, \bibinfo{pages}{1--9}.
\newblock


\bibitem[Pennycook et~al\mbox{.}(2019)]%
        {pennycook2019implications}
\bibfield{author}{\bibinfo{person}{Simon~J Pennycook}, \bibinfo{person}{Jason~D Sewall}, {and} \bibinfo{person}{Victor~W Lee}.} \bibinfo{year}{2019}\natexlab{}.
\newblock \showarticletitle{Implications of a metric for performance portability}.
\newblock \bibinfo{journal}{\emph{Future Generation Computer Systems}}  \bibinfo{volume}{92} (\bibinfo{year}{2019}), \bibinfo{pages}{947--958}.
\newblock


\bibitem[Pilz et~al\mbox{.}(2025)]%
        {pilz2025trendsaisupercomputers}
\bibfield{author}{\bibinfo{person}{Konstantin~F. Pilz}, \bibinfo{person}{James Sanders}, \bibinfo{person}{Robi Rahman}, {and} \bibinfo{person}{Lennart Heim}.} \bibinfo{year}{2025}\natexlab{}.
\newblock \bibinfo{title}{Trends in AI Supercomputers}.
\newblock
\newblock
\showeprint[arxiv]{2504.16026}~[cs.CY]
\urldef\tempurl%
\url{https://arxiv.org/abs/2504.16026}
\showURL{%
\tempurl}


\bibitem[Piñeiro and Pichel(2026)]%
        {PINEIRO2026108035}
\bibfield{author}{\bibinfo{person}{César Piñeiro} {and} \bibinfo{person}{Juan~C. Pichel}.} \bibinfo{year}{2026}\natexlab{}.
\newblock \showarticletitle{OMP4Py: A pure Python implementation of openMP}.
\newblock \bibinfo{journal}{\emph{Future Generation Computer Systems}}  \bibinfo{volume}{175} (\bibinfo{year}{2026}), \bibinfo{pages}{108035}.
\newblock
\showISSN{0167-739X}
\urldef\tempurl%
\url{https://doi.org/10.1016/j.future.2025.108035}
\showDOI{\tempurl}


\bibitem[Poenaru et~al\mbox{.}(2021)]%
        {poenaru2021performance}
\bibfield{author}{\bibinfo{person}{Andrei Poenaru}, \bibinfo{person}{Wei-Chen Lin}, {and} \bibinfo{person}{Simon McIntosh-Smith}.} \bibinfo{year}{2021}\natexlab{}.
\newblock \showarticletitle{A performance analysis of modern parallel programming models using a compute-bound application}. In \bibinfo{booktitle}{\emph{International Conference on High Performance Computing}}. Springer, \bibinfo{pages}{332--350}.
\newblock


\bibitem[Raihan et~al\mbox{.}(2025)]%
        {raihan-etal-2025-mojobench}
\bibfield{author}{\bibinfo{person}{Nishat Raihan}, \bibinfo{person}{Joanna C.~S. Santos}, {and} \bibinfo{person}{Marcos Zampieri}.} \bibinfo{year}{2025}\natexlab{}.
\newblock \showarticletitle{{M}ojo{B}ench: Language Modeling and Benchmarks for Mojo}. In \bibinfo{booktitle}{\emph{Findings of the Association for Computational Linguistics: NAACL 2025}}, \bibfield{editor}{\bibinfo{person}{Luis Chiruzzo}, \bibinfo{person}{Alan Ritter}, {and} \bibinfo{person}{Lu~Wang}} (Eds.). \bibinfo{publisher}{Association for Computational Linguistics}, \bibinfo{address}{Albuquerque, New Mexico}, \bibinfo{pages}{4109--4128}.
\newblock
\showISBNx{979-8-89176-195-7}
\urldef\tempurl%
\url{https://doi.org/10.18653/v1/2025.findings-naacl.230}
\showDOI{\tempurl}


\bibitem[Samaroo et~al\mbox{.}(2023)]%
        {julian_samaroo_2023_10040461}
\bibfield{author}{\bibinfo{person}{Julian Samaroo} {et~al\mbox{.}}} \bibinfo{year}{2023}\natexlab{}.
\newblock \bibinfo{booktitle}{\emph{JuliaGPU/AMDGPU.jl: v0.7.3}}.
\newblock
\urldef\tempurl%
\url{https://doi.org/10.5281/zenodo.10040461}
\showDOI{\tempurl}


\bibitem[Saraswat et~al\mbox{.}(2007)]%
        {x10}
\bibfield{author}{\bibinfo{person}{Vijay~A. Saraswat}, \bibinfo{person}{Vivek Sarkar}, {and} \bibinfo{person}{Christoph von Praun}.} \bibinfo{year}{2007}\natexlab{}.
\newblock \showarticletitle{X10: {C}oncurrent {P}rogramming for {M}odern {A}rchitectures}. In \bibinfo{booktitle}{\emph{Proceedings of the 12th ACM SIGPLAN Symposium on Principles and Practice of Parallel Programming}} (San Jose, California, USA) \emph{(\bibinfo{series}{PPoPP '07})}. \bibinfo{publisher}{Association for Computing Machinery}, \bibinfo{address}{New York, NY, USA}, \bibinfo{pages}{271}.
\newblock
\showISBNx{9781595936028}
\urldef\tempurl%
\url{https://doi.org/10.1145/1229428.1229483}
\showDOI{\tempurl}


\bibitem[Stone et~al\mbox{.}(2010)]%
        {stone2010opencl}
\bibfield{author}{\bibinfo{person}{John~E Stone} {et~al\mbox{.}}} \bibinfo{year}{2010}\natexlab{}.
\newblock \showarticletitle{OpenCL: A parallel programming standard for heterogeneous computing systems}.
\newblock \bibinfo{journal}{\emph{Computing in science \& engineering}} \bibinfo{volume}{12}, \bibinfo{number}{3} (\bibinfo{year}{2010}), \bibinfo{pages}{66}.
\newblock


\bibitem[Trott et~al\mbox{.}(2022)]%
        {Kokkos3}
\bibfield{author}{\bibinfo{person}{Christian~R. Trott} {et~al\mbox{.}}} \bibinfo{year}{2022}\natexlab{}.
\newblock \showarticletitle{Kokkos 3: Programming Model Extensions for the Exascale Era}.
\newblock \bibinfo{journal}{\emph{IEEE Transactions on Parallel and Distributed Systems}} \bibinfo{volume}{33}, \bibinfo{number}{4} (\bibinfo{year}{2022}), \bibinfo{pages}{805--817}.
\newblock
\urldef\tempurl%
\url{https://doi.org/10.1109/TPDS.2021.3097283}
\showDOI{\tempurl}


\bibitem[Valero-Lara et~al\mbox{.}(2024)]%
        {10820713}
\bibfield{author}{\bibinfo{person}{Pedro Valero-Lara} {et~al\mbox{.}}} \bibinfo{year}{2024}\natexlab{}.
\newblock \showarticletitle{JACC: Leveraging HPC Meta-Programming and Performance Portability with the Just-in-Time and LLVM-based Julia Language}. In \bibinfo{booktitle}{\emph{SC24-W: Workshops of the International Conference for High Performance Computing, Networking, Storage and Analysis}}. \bibinfo{pages}{1955--1966}.
\newblock
\urldef\tempurl%
\url{https://doi.org/10.1109/SCW63240.2024.00245}
\showDOI{\tempurl}


\bibitem[Van~Rossum et~al\mbox{.}(2007)]%
        {van2007python}
\bibfield{author}{\bibinfo{person}{Guido Van~Rossum} {et~al\mbox{.}}} \bibinfo{year}{2007}\natexlab{}.
\newblock \showarticletitle{Python Programming Language.}. In \bibinfo{booktitle}{\emph{USENIX annual technical conference}}, Vol.~\bibinfo{volume}{41}. Santa Clara, CA, \bibinfo{pages}{1--36}.
\newblock


\bibitem[Vetter et~al\mbox{.}(2018)]%
        {osti_1473756}
\bibfield{author}{\bibinfo{person}{Jeffrey~S. Vetter}, \bibinfo{person}{Ron Brightwell}, \bibinfo{person}{Maya Gokhale}, \bibinfo{person}{Pat McCormick}, \bibinfo{person}{Rob Ross}, \bibinfo{person}{John Shalf}, \bibinfo{person}{Katie Antypas}, \bibinfo{person}{David Donofrio}, \bibinfo{person}{Travis Humble}, \bibinfo{person}{Catherine Schuman}, \bibinfo{person}{Brian Van~Essen}, \bibinfo{person}{Shinjae Yoo}, \bibinfo{person}{Alex Aiken}, \bibinfo{person}{David Bernholdt}, \bibinfo{person}{Suren Byna}, \bibinfo{person}{Kirk Cameron}, \bibinfo{person}{Frank Cappello}, \bibinfo{person}{Barbara Chapman}, \bibinfo{person}{Andrew Chien}, \bibinfo{person}{Mary Hall}, \bibinfo{person}{Rebecca Hartman-Baker}, \bibinfo{person}{Zhiling Lan}, \bibinfo{person}{Michael Lang}, \bibinfo{person}{John Leidel}, \bibinfo{person}{Sherry Li}, \bibinfo{person}{Robert Lucas}, \bibinfo{person}{John Mellor-Crummey}, \bibinfo{person}{Paul Peltz~Jr.}, \bibinfo{person}{Thomas Peterka}, \bibinfo{person}{Michelle Strout}, {and}
  \bibinfo{person}{Jeremiah Wilke}.} \bibinfo{year}{2018}\natexlab{}.
\newblock \showarticletitle{{Extreme {H}eterogeneity 2018 - {P}roductive {C}omputational {S}cience in the {E}ra of {E}xtreme {H}eterogeneity: {R}eport for {DOE} {ASCR} {W}orkshop on {E}xtreme {H}eterogeneity}}.
\newblock  (\bibinfo{date}{12} \bibinfo{year}{2018}).
\newblock
\urldef\tempurl%
\url{https://doi.org/10.2172/1473756}
\showDOI{\tempurl}


\bibitem[Wienke et~al\mbox{.}(2012)]%
        {wienke2012openacc}
\bibfield{author}{\bibinfo{person}{Sandra Wienke} {et~al\mbox{.}}} \bibinfo{year}{2012}\natexlab{}.
\newblock \showarticletitle{OpenACC—first experiences with real-world applications}. In \bibinfo{booktitle}{\emph{Euro-Par 2012 Parallel Processing: 18th International Conference}}. Springer, \bibinfo{pages}{859--870}.
\newblock


\end{thebibliography}

\appendix

\twocolumn[%
{\begin{center}
\Huge
Appendix: Artifact Description/Artifact Evaluation        
\end{center}}
]

\appendixAD

\section{Overview of Contributions and Artifacts}

\subsection{Paper's Main Contributions}

The main goal of the paper is to understand the recently added performance portability capabilities of the MLIR (Multi-Level Intermediate Representation)-based Mojo on modern NVIDIA and AMD (since June 2025) GPUs. Mojo is a novel programming language that aims to close performance gaps using an interoperable compile-time superset that is interoperable at run-time with the full Python ecosystem. All the codes used for this study are provided on a single repository: \url{https://github.com/tdehoff/Mojo-workloads} hosted on GitHub with a permissive MIT open-source software license.

\begin{description}
\item[$C_1$] Understand the performance of GPU-portable Mojo code against CUDA and HIP targeting widely-used kernels across science domains: Seven-point stencil, BabelStream, Hartree-Fock and miniBUDE.
\item[$C_2$] Understand observed performance differences (over and under-performance) on NVIDIA H100 and AMD MI300A GPUs at a lower-level using NVIDIA's Nsight profiling tools.
\item[$C_3$] Develop, document and provide readily-available open-source Mojo ports of relevant these science kernels to the community for experimentation with this novel language.
\end{description}

\subsection{Computational Artifacts}


\begin{description}
\item[$A_1$] New Mojo ports, and existing CUDA and HIP codes: \\ 
\url{https://github.com/tdehoff/Mojo-workloads}
\end{description}


\begin{center}
\begin{tabular}{rll}
\toprule
Artifact ID  &  Contributions &  Related \\
             &  Supported     &  Paper Elements \\
\midrule
$A_1$   &  $C_1$ and $C_3$ & Tables 2-5 \\
        &        & Figure 2-7\\
\bottomrule
\end{tabular}
\end{center}

\section{Artifact Identification}


\newartifact

\artrel

All the provided Mojo codes have been developed to generate the results on this paper. The listed $A_1$ repository contains the following subdirectories, each containing a relevant proxy applications used in our evaluations and will produce the attached relevant metric listed as follows:

\begin{itemize}
    \item seven-point stencil (memory bandwidth)
    \item babelstream (memory bandwidth)
    \item miniBude (compute-bound)
     \item hartree-fock (kernel wall-clock time)
\end{itemize}

Each of these directories contains subdirectories with the portable Mojo, and the vendor-specific CUDA, and HIP implementations used for the performance study. Hence Figures 2 to 7 will be straight-forward plots of the resulting runs containing 1000 iterations to capture computational variability.

\artexp

The expected results are the performance metrics, in Eqs. ~\ref{eqn:7p_bandwidth_eff} for seven-point stencil,~\ref{eqn:babelstream_bandwidth_eff} for BabelStream, \ref{eqn:minibude_gflops} for miniBUDE, and wall-clock time for Hartree-Fock, that are part of the output when running the listed codes. As shown in the Figures 2-6 and Tables 2-4, we expect that Mojo codes will be on-par with the HIP versions on AMD GPUs. Whereas differences reported with CUDA code on NVIDIA GPUs should be reproducible. Table 2-4 results require the use of the NVIDIA Nsight CLI (ncu) profiling tool. We profile using the ncu command line as detailed in the next section.

\arttime


Each of the runs can be executed in less than 1 minute per case. Users can select the number of iterations to test the kernels. 

{\it Artifact Setup:} all codes are hosted publicly on GitHub, or just downloading them from Zenodo. CUDA and HIP C\texttt{++} codes use Makefiles for the appropriate target, e.g. \texttt{make cuda}. Compilation of all codes should be take 1-5 seconds as these are proxy apps.

{\it Artifact Execution:}

\begin{itemize}
    \item Seven-point stencil: 5 s 
    \item BabelStream: 5 s
    \item miniBude: 5 s
    \item Hartree-Fock: 5s, he1024 case take 124 s on H100
\end{itemize}

{\it Artifact Analysis:}
We provide Python plotting scripts that take the generated code output data as inputs. It captures performance numbers as text and generates Figures 2-3 shown in the paper. It should take less than 2 s to generate each plot.

\paragraph{Profiling} we use NVIDIA Nsight CLI (ncu) that ships as part of NVIDIA HPC SDK (nvhpc). The following command is a straight-forward use-case the same code at a smaller number of iterations (e.g., 10). \\

\begin{lstlisting}[language=Markdown, basicstyle=\small,linewidth=0.99\linewidth,xleftmargin=.01\linewidth]
$ ncu -f -o output --set roofline ./babelstream 
\end{lstlisting}

Mojo code need to be pre-compiled (e.g. \texttt{program}) by running the following commands: 

\begin{lstlisting}[language=Markdown, basicstyle=\small]
$ pixi shell 
(mojo) $ mojo build babelstream.mojo
(mojo) $ ./babelstream
\end{lstlisting}

\artin

\artinpart{Hardware}

The GPUs used in this study are listed in Table~\ref{tab:HardwareAppendix}.

\begin{table}[h]
\small
\centering
\resizebox{\linewidth}{!}{%
\begin{tabular}{@{}crrr@{}}
\toprule
GPU - Memory   & \multicolumn{3}{c}{Theoretical Peaks}  \\
       &  Bandwidth      & FP32     & FP64      \\
       &  GB/s           & TFLOPS/s & TFLOPS/s  \\ 
\midrule
NVIDIA H100 NVL - 94\,GB  &  3,900  &  60.0 & 30.0  \\
AMD MI300A - 128 GB HBM3  &  5,300  & 122.6 & 61.3  \\
\bottomrule
\end{tabular}%
}
\caption{GPU hardware used in this study}
\label{tab:HardwareAppendix}
\end{table}

\artinpart{Software} 

We use a single self-contained repository for all the codes used in this study:
\begin{itemize}
    \item Mojo-workloads with $A_1$ codes: \url{https://github.com/tdehoff/Mojo-workloads}
    
\end{itemize}

\artinpart{Datasets / Inputs}
No special datasets are used as inputs. Hartree-Fock codes use the original dataset files provided by the original Fortran proxy app in \url{https://github.com/gdfletcher/basic-hf-proxy/tree/main/tests}. These files are provided under the hartree-fock subdirectory.
Inputs can be directly modified on each file (e.g. problem size and types) to match the Figures in the paper. It should be very intuitive as each kernel only depend on a few problems sizes, types and number of iterations. 


\artinpart{Installation and Deployment}


Code subdirectories in $A_1$ can be considered stand-alone executables. They need the following Mojo and NVIDIA CUDA and AMD ROCM compiler versions:

\begin{itemize}
\item Mojo: Mojo 25.5.0.dev2025062505 \url{https://docs.modular.com/mojo/manual/gpu/intro-tutorial/}
    \item C\texttt{++} CUDA codes require NVIDIA HPC SDK (nvhpc) v24.9 \url{https://developer.nvidia.com/hpc-sdk-releases}
    \item C\texttt{++} HIP codes require AMD ROCM v6.4.0 \url{https://github.com/ROCm/ROCm/tree/rocm-6.4.0}
\end{itemize}

\artcomp

All the results can be easily reproducible. We summarize the typical runs and configuration parameters for each Mojo port of the following proxy applications:

\begin{itemize}
\item seven-point stencil
\item BabelStream
\item miniBUDE
\item Hartree-Fock (input data in test directory)
\end{itemize}

Modifying configuration parameters is straightforward and they represent job characteristics related to problem size, type (where appropriate), and GPU threads-per-block (TBSize). Vendor-specific CUDA and HIP implementations are part of the directories structure and each proxy and model combination has its own \texttt{makefile} that contains the relevant compiler flags. The following are the common parameters and the instructions for building and running the novel Mojo implementations.

\paragraph{Seven-point stencil} uses the following parameters that can be edited in the source code. 

\begin{itemize}
    \item Problem sizes of L = 512 and 1024
    \item TBSize = (1024, 1, 1) or (512, 1, 1)
    \item Iterations, num\_iter = 1000
    \item Precision = \texttt{Float32} or \texttt{Float64}
\end{itemize}

\begin{lstlisting}[caption="Seven-point stencil run options",language=Markdown, basicstyle=\small,linewidth=0.9\linewidth,xleftmargin=.1\linewidth]
JIT run
$ pixi run mojo laplacian.mojo --csv
AOT run
$ pixi shell
(mojo)$ mojo build laplacian.mojo
(mojo)$ ./laplacian --csv
\end{lstlisting}

\paragraph{BabelStream} the main parameter is the problem size. In our results, we use a size of $2^{25} = 33,554,432$ allowing us to saturate GPU device memory. Since we use a 1D computational grid, we only set the first dimension to a fixed value of 1,024. BabelStream will run all the kernels in the benchmark: Copy, Mul, Add, Triad and Dot.

\begin{itemize}
    \item Problem sizes of L = pow(2,15)
    \item TBSize = 1024 (1024, 1, 1)
    \item Iterations, num\_iter = 1000
    \item Precision = \texttt{Float32} or \texttt{Float64}
\end{itemize}

\begin{lstlisting}[caption="miniBUDE run options", language=Markdown, basicstyle=\small,linewidth=0.9\linewidth,xleftmargin=.1\linewidth]
JIT run
$ pixi run mojo babelstream.mojo --csv
AOT run
$ pixi shell
(mojo)$ mojo build babelstream.mojo
(mojo)$ ./babelstream --csv
\end{lstlisting}

\paragraph{miniBUDE} we limit the scope to the bm1 benchmark. The main parameters are the poses per work-item (PPWI), and (ii) work-group (wg) size, all other parameters remain constant including the number of poses set to a large value of 65,536. We use sizes of PPWI that are a power of 2 from 1 to 128, while wg is set for 2 values of 8 and 64. Floating Precision is fixed due to the nature of the kernel. 

\begin{itemize}
    \item Problem sizes PPWI = {1,2,4,8,16,32,64,128}, wg = {8,64}
    \item num\_poses = 65,536
    \item TBSize = 1024 (1024, 1, 1)
    \item Iterations, num\_iter = 1000
\end{itemize}

\begin{lstlisting}[caption="miniBUDE run options", language=Markdown, basicstyle=\small,linewidth=0.9\linewidth,xleftmargin=.1\linewidth]
JIT run
$ pixi run mojo minibude.mojo
AOT run
$ pixi shell
(mojo)$ mojo build minibude.mojo --csv
(mojo)$ ./minibude --csv
\end{lstlisting}

\paragraph{Hartree-Fock} runs are set to a fixed size of atoms modeling a system of Helium atoms. The datasets needed for each run are included inside the \texttt{test} directory and each file needs to be passed as an argument. We ran successfully up to a number of atoms of 256, while we observe abnormal behaviors in terms of performance for the 512 and 1024 cases. All other parameters are kept constant. The parameters natoms and ngauss must be modified in the source code to match the corresponding file in the \texttt{test} directory.

\begin{itemize}
    \item Problem sizes natoms = {64, 128, 256, 1024} ngauss = 3 or 6 (1024 only)
    \item Iterations, num\_iter = 1000
\end{itemize}

\begin{lstlisting}[caption="Hartree-Fock run options for 256 atoms test", language=Markdown, basicstyle=\small,linewidth=1\linewidth,xleftmargin=0.005\linewidth]
JIT run
$ pixi run mojo hartree-fock.mojo ../tests/he256
AOT run
$ pixi shell
(mojo)$ mojo build hartree-fock.mojo
(mojo)$ ./hartree-fock ../tests/he256 --csv
\end{lstlisting}




\appendixAE

\arteval{1}
\artin

The single source code repository in \url{https://github.com/tdehoff/Mojo-workloads}, for Mojo and vendor-specific CUDA and HIP versions and scripts for plotting the results, is structured as follows:

<science-workload>/
\begin{itemize}
      \item Mojo 
      \item cuda 
      \item hip
      \item plotting
\end{itemize}

where <science-workload> are: \texttt{7-point-stencil}, \texttt{babelstream}, \texttt{hartree-fock}, and \texttt{miniBUDE} subdirectories. 

Thus each Mojo, CUDA and HIP version runs independently, respectively. CUDA and HIP versions posses a \texttt{makefile}, whereas Mojo versions have a single mojo source code and pixi toml files for reproducibility. 
Plotting files are simple Python's \texttt{pandas}, \texttt{matplotlib}, and \texttt{seaborn} scripts to generate Figs, 3, 4 and 6 for each application, except Hartree-Fock for which wall-clock experiments' runs results are presented in Table 4.  

\subsection{Compiling libraries and code}

Each Mojo, CUDA, HIP rely on standard installations of the language, and compiler toolchains widely available for any Linux distribution system: \\

Mojo (both NVIDIA and AMD GPUs): \url{https://docs.modular.com/mojo/manual/gpu/intro-tutorial/#1-create-a-mojo-project}
\begin{itemize}
\item pixi 0.48.2
\item Mojo 25.5.0.dev2025070105
\end{itemize}

CUDA: \url{https://developer.nvidia.com/hpc-sdk-releases}
\begin{itemize}
    \item nvhpc v24.9
    \item cuda 12.6
    \item gcc/g++ 13.3.0
\end{itemize}

HIP: \url{https://rocm.docs.amd.com/projects/install-on-linux/en/latest/}
\begin{itemize}
    \item rocm v6.4.0
    \item gcc/g++ 13.3.0
\end{itemize}

To compile Mojo code (e.g. Seven-point stencil's laplacian.mojo file):

\begin{lstlisting}[caption="Seven-point stencil run options",language=Markdown, basicstyle=\small,linewidth=0.9\linewidth,xleftmargin=.1\linewidth]
JIT run
$ pixi run mojo laplacian.mojo --csv
AOT run
$ pixi shell
(mojo)$ mojo build laplacian.mojo
(mojo)$ ./laplacian --csv
\end{lstlisting}

To compile CUDA or HIP code execute the corresponding Makefile (using \texttt{make}) inside the \texttt{<science-workload>/cuda} and the \texttt{<science-workload>/hip} subdirectories. The following code presents the Makefile for BabelStream using CUDA on a NVIDIA H100 which will generate the corresponsing executable:

\begin{lstlisting}[caption="BabelStream CUDA makefile",language=Markdown, basicstyle=\small,linewidth=0.9\linewidth,xleftmargin=.1\linewidth]
$ make
Makefile:

CC = nvcc
FLAGS=-forward-unknown-to-host-compiler -arch=sm_90  -DNDEBUG -O3 -march=native -std=c++11

SRCS = main.cpp CUDAStream.cu
HEADERS = Stream.h CUDAStream.h
TARGET = babelstream_cuda

all: $(TARGET)

$(TARGET): $(SRCS) $(HEADERS)
	$(CC) $(FLAGS) -o $@ $(SRCS)

clean:
	rm -f $(TARGET)

\end{lstlisting}

\subsection{Workflow description}

Example using the seven-point stencil case as a reference to generate Figs.~\ref{fig:7p-CUDA} and ~\ref{fig:7p-HIP}: \\

\noindent 1. Clone the GitHub repo: 
\\
\\
\texttt{\$ git clone https://github.com/tdehoff/Mojo-workloads}
\\
\\
2. Run the Mojo code with csv output (e.g. JIT run) \\
\\
\texttt{cd 7-point-stencil/Mojo} \\
\texttt{\$ pixi run mojo laplacian.mojo --csv} \\
\\
\\
3. Run the CUDA (or HIP) code \\
\\
\texttt{cd ../cuda}
set \texttt{csv\_output} to \texttt{true} in the code.
\texttt{\$ make}\\
\texttt{\$ ./laplacian\_kernel} 
\\
\\
4. Merge outputs from Step 2 and 3 into a single csv file and run the plotting scripts to generate Fig. 3: \\
\\
\texttt{python ./plot\_NVIDIA.py} \\
- or - \\
\texttt{python ./plot\_AMD.py} \\
\\
Examples for CSV files used in this study are provided in the \texttt{plotting} subdirectories. 

Repeat the same process for BabelStream and miniBUDE.







\end{document}